\documentclass[
 reprint,
 amsmath,amssymb,
 aps,
 nofootinbib
]{revtex4-2}

\usepackage{graphicx}
\usepackage{dcolumn}
\usepackage{bm}
\usepackage{xcolor}
\usepackage{hyperref}
\usepackage[normalem]{ulem}
\usepackage{xtab,afterpage,longtable}

\hypersetup{
colorlinks=true,
citecolor=blue,
linkcolor=blue,
urlcolor=blue}

\begin{document}

\preprint{APS/123-QED}

\title{Super-knee cosmic rays from interacting supernovae}

\author{Nick Ekanger$^{1,2}$}
\email{ekangernj@astr.tohoku.ac.jp}
\author{Shigeo S. Kimura$^{1,2}$}
\email{shigeo@astr.tohoku.ac.jp}
\author{Kazumi Kashiyama$^{2}$}
\email{kashiyama@astr.tohoku.ac.jp}
\affiliation{${}^1$Frontier Research Institute for Interdisciplinary Sciences, Tohoku University, Sendai 980-8578, Japan}
\affiliation{${}^2$Astronomical Institute, Graduate School of Science, Tohoku University, Sendai 980-8578, Japan}

\date{\today}

\begin{abstract}
There is increasing evidence that, in the very late phase of stellar evolution before core collapse, massive stars have winds with large mass loss rates that give rise to a dense circumstellar medium (CSM) surrounding the progenitor star. After core collapse, a shock wave forms when the supernova ejecta interacts with this CSM. In such an interaction, the nuclei in the CSM can undergo diffusive shock acceleration and reach very high energies. We consider such a model, which includes magnetic field amplification from the non-resonant streaming instability, enhancement to the abundance of heavy-ions, and composition-dependent acceleration. Applying this to several supernova subclasses, we find that IIn supernovae can supply a dominant fraction of the observed super-knee cosmic-ray (CR) flux from $\sim{\rm few}\times10^{15}\,{\rm eV}$ to $\sim{\rm few}\times10^{17}\,{\rm eV}$ and is consistent with recent LHAASO measurements above the CR knee. This systematic model also explains the increasingly heavy nuclear composition in this energy range.
\end{abstract}

\maketitle

\section{\label{sec:intro}Introduction}

Cosmic-rays (CRs) at high energies are mainly charged nuclei, so they are deflected by intervening magnetic fields on their way to Earth. The source populations of high energy CRs, then, remain mysterious. The CR flux is well measured as a function of energy by a combination of direct-detection and air-shower experiments (see, e.g., \cite{Nagano00,KASCADE:2005ynk,Hillas05,Kotera11,Kachelriess:2019oqu}) and can be well described by a set of broken power-law functions. The CR `knee' - the first important change in power law index - occurs around $\sim{\rm few}\times10^{15}\,{\rm eV}$. Below this energy, in the Sedov-Taylor phase thousands of years after the supernova (SN), supernova remnants (SNRs) are the primary source of galactic CRs as they sweep up the interstellar medium (see \cite{Hillas05,Bell13,Blasi:2013rva,Giuliani:2024fpq}), but are limited in the maximum energy they can achieve \cite{Thoudam:2016syr,Diesing:2023ldd}. There is another spectral feature - a break in the power-law index - around $\sim10^{17}\,{\rm eV}$ deemed the `second knee.'

Above the knee energies, there are third and fourth features marking additional changes in the power-law index (the `ankle' at $\sim{\rm few}\times10^{18}\,{\rm eV}$) and a cutoff to the spectrum where the flux decreases significantly (consistent with ultra-high energy CRs interacting with cosmic radiation backgrounds, known as the `GZK' effect \cite{Greisen:1966jv,1966JETPL...4...78Z}). CRs above the ankle are expected to come from distant, extragalactic sources because there are few feasible galactic sources able to accelerate nuclei to such energies and a source at short distances would be traceable, despite magnetic fields (discussed in, e.g., \cite{Hillas05}).

Between the first knee and the ankle, however, sources of CRs are still unknown. These sources can be galactic or nearby, since CRs at this energy can diffuse in galactic magnetic fields. Recent air-shower experiments, including the Pierre Auger Observatory (PAO) \cite{PAO2015}, Telescope Array (TA) \cite{Abu_Zayyad_2013}, IceTop \cite{IceTop13}, and LHAASO \cite{Cao:2010zz,He18}, have measured the energy flux in this energy region well. These experiments have also observed another important clue for demystifying CR sites: composition of nuclei. It seems that the composition tends to change in coincidence with the spectral features; loosely speaking, it is light at the knee, heavy at the second knee, light again at the ankle, and heavy at the cutoff (see, e.g., \cite{Morello06,Abu-Zayyad:2018btv,Falalaki:2024gst}). This suggests that there may be distinct populations of sources at spectral features. 

In this energy region between the knee and the ankle, the theoretical class of CR accelerators is named `PeVatrons.' There are many proposed PeVatrons so far, including stellar winds \cite{Cesarsky83,Aharonian:2018oau}, pulsars \cite{Bednarek:2001av,Giller02,Bednarek:2004wp,Ohira:2017bxa}, young massive stellar clusters \cite{Bykov:2020zqf,Morlino1:2021zwu,Vieu:2022mas,Vieu:2022wsc}, hypernovae \cite{Sveshnikova03}, isolated black holes \cite{Ioka:2016bil,Kimura:2024izc}, interacting supernovae (ISN) \cite{Murase:2013kda,Inoue:2021bjx,Brose:2025npd}, and microquasars \cite{Zhang:2025tew}. If CRs are accelerated beyond $\gtrsim10^{16}\,{\rm eV}$, they may be referred to as `Super-PeVatrons,' which may be the case for super-accreting X-ray binaries \cite{Wang:2025yqy}, for example. All or some of these may contribute to the CR spectrum and be able to explain the composition as well. 

The key to understanding the dominant contributors may be through a multimessenger lens. LHAASO has discovered many high energy gamma-rays above $\sim100\,{\rm TeV}$ \cite{LHAASO:2021gok,LHAASO:2021plb,LHAASO:2021crt,LHAASO:2023rpg}, another important piece of evidence for PeV acceleration. However, gamma-rays could be produced through leptonic processes, namely Inverse Compton scattering \cite{Cardillo:2023hbb}. Pulsar wind nebulae are an important class of potential leptonic accelerators, as observed by HAWC \cite{HAWC:2019xhp,HAWC:2019tcx}, H.E.S.S. \cite{HESS:2018usw}, and the Tibet AS array \cite{TibetASg:2019ivi}. Then, the detection of neutrinos would be a smoking-gun of hadronic particle acceleration and help to explain the sources of the CR spectrum. Recently, IceCube has detected neutrinos from the Galactic plane \cite{IceCube:2023ame}, but whether these neutrinos are from individual, unresolved sources, from diffuse emission from CR interactions with the ISM, or from both is undetermined \cite{Fang:2024fyd}. Studies considering the galactic emission of gamma-rays also hint at this connection, but are not yet definitive \cite{Fang:2024fyd,Li:2025ank}. SNRs are a well motivated candidate for this multimessenger emission \cite{Gagliardini:2024een,Simon:2025axa}. For example, the unidentified LHAASO source J2108+5157 could be one such example of gamma-rays from an old, supernova-like explosion \cite{delaFuente:2023rqo,Mitchell:2023qex} that underwent CR acceleration within the Galaxy. Future observations with the KM3NeT observatory will help clear up the fuzzy galactic emission picture \cite{KM3NeT:2018wnd,Gagliardini:2024een}.

In this work, we also examine ISNe as potential Super-PeVatron sources. In these systems, ejecta interacts with pre-supernova mass-loss known as the circumstellar material (CSM) in the tens to hundreds of days after the SN, long before the Sedov-Taylor phase. The faster SN ejecta forms a shock and, as the shock propagates, nuclei can be accelerated through diffusive shock acceleration (DSA) (see, e.g., \cite{Drury83}). The flux and energy of the accelerated nuclei depends on physical parameters of the SN and the CSM. In recent years, the mass loss rate inferred from ISNe has generally increased (see, e.g., estimates like Ref.~\cite{Fassia:2000ee}). Additionally, Ibn \cite{Pastorello:2008vy} and Icn \cite{Gal-Yam:2021rdf} are somewhat newly classified ISN types. Thus, a denser CSM than previously expected seems to be a prevalent feature among many supernova types, which could be an ideal environment for particle acceleration.

Here, we build on a phenomenological model of the CSM-ejecta shock interaction to investigate cosmic-ray production in several ISN types. In order to have efficient acceleration, the shock must be sufficiently optically thin, i.e., it transitions from a radiation-mediated to a collisionless shock when the density decreases. The maximum energy of nuclei is determined by competing destruction and acceleration processes, which depend on their masses and atomic numbers. We consider an escape-limited model, where only the highest-energy nuclei will escape from the system from each event. We then account for the propagation of escaping cosmic-rays as they travel through the interstellar medium. Finally, we estimate the total contribution of various ISN types to the $\gtrsim {\rm PeV}$ CR flux and composition measured on Earth.

This paper is organized as follows. In Section~\ref{sec:overview}, we describe the overall physical picture of ISNe to motivate the following sections. In Section~\ref{sec:SNe}, we describe the evolution of the shock and the types of ISNe we consider. In Section~\ref{sec:acceleration} we describe the competing processes that determine the cosmic-ray maximum energy and composition. In Section~\ref{sec:fluxcomp} we calculate the resulting CR spectrum and predict the observable flux and composition with our models, which we compare to observed data. In Section~\ref{sec:disc} we discuss caveats and extensions to our work. In Section~\ref{sec:summary} we summarize our results. Finally, in Appendix~\ref{sec:icetopresults}, we compare our models to additional data from the IceTop experiment.

\section{\label{sec:overview}Model Overview}

In this section we overview the physical picture of interacting supernovae. Here, we also estimate the particle acceleration results and the expected flux of CRs with simple analytical estimates, before describing the detailed modeling in later sections.

Before the core-collapse of ISNe, these massive progenitors undergo periods of mass-loss which results in a low-velocity, massive CSM surrounding the star. After core-collapse, the SN ejecta expands into the slower CSM, which forms a shock. 

Initially, this shock may be radiation-mediated and a poor environment for accelerating particles. As the shock expands, the optical depth decreases and photons are able to escape as the shock becomes collisionless. Nuclei, trapped by the magnetic field in this region, gain energy as they bounce across the shock boundary. The abundance of nuclei injected depends on their effective charge, so their ionization degree when they are injected into the shock must be evaluated. Their energy gain also depends on the strength of the magnetic field. This determines the Larmor radius and, thus, the length and timescale of acceleration across the shock boundary. 

We can estimate the maximum energy these nuclei can reach in the following way. We first assert that the maximum energy, $E_{\rm max}$, can approximately be set by this Larmor radius lengthscale, i.e., $r_L=E_{\rm max}/(ZeB_w)$, where $Z$ is the nucleus charge and $B_w$ is the magnetic field strength of the wind. We can assume that the magnetic field energy density is proportional to the kinetic energy density of the wind: $B_w^2/(8\pi)\approx\epsilon_B\rho_wV_w^2/2$, where $\epsilon_B\sim10^{-3}$ is the fraction of the magnetic field density compared to the kinetic energy density, and $\rho_w$ and $V_w$ are the density and velocity of the wind, respectively. This wind density is given by $\rho_w=\dot{M}/(4\pi R_0^2V_w)$, where $\dot{M}$ is the wind mass-loss rate and $R_0$ is the characteristic lengthscale of the shock, which we can approximate as $r_L$. Thus, $E_{\rm max}\approx Ze(\epsilon_B\dot{M}V_w)^{1/2}\approx 5\times10^{16}\,{\rm eV}$\footnote{Note that this estimation ignores a factor of $V_{\rm sh}/c$ for the acceleration of nonrelativistic particles, magnetic field amplification, and diffusion, but accounting for these yields a similar result and is properly accounted for in later sections.} for $Z=2$, $\dot{M}\sim10^{-2}\,{\rm M_{\odot}~yr^{-1}}$, and $V_w\sim100\,{\rm km~s^{-1}}$.

In reality, the maximum energy can be limited by destruction processes in this dense environment, like spallation and photodisintegration, by expansion, and by the escape of nuclei. This competition of processes determines the true maximum energy and the time when they can escape as cosmic rays. This time is determined by when the escape timescale is the fastest process, which we label as $t_{\rm crit}$. In this work, we consider an escape-limited model (see \cite{Zhang:2017moz}), where only CRs near the maximum energy escape from the system. We can then briefly estimate the flux of PeV CRs from ISNe.

We can do this by comparing the ratio of the luminosity required to provide the observed CR flux at $\sim1\,{\rm GeV}$, which primarily comes from SNRs of typical supernovae, and at $E_{\rm max}\sim50\,{\rm PeV}$, which may come from ISNe. This luminosity ratio, $L_{\rm CR}|_{\rm GeV}/L_{\rm CR}|_{50~\rm PeV}\sim10^3$ (see Sec. 4.1 of Ref.~\cite{Ioka:2016bil}). Note that we do not compare the observed flux at these energies because the observed flux accounts for propagation effects. We can approximate this ratio by assuming $L_{\rm CR}\propto\epsilon_{\rm CR}MV^2\mathcal{R}/2$, for the mass swept up $M$, shock velocity $V$, and rate $\mathcal{R}$. Additionally, in the escape-limited case for ISNe, we divide by a bolometric correction ($\ln(E_{\rm max}/E_{\rm min})\sim10$ for an $E_{\rm min}\sim1\,{\rm GeV}$, see Sec.~\ref{sec:fluxcomp}). However, the SNR case does not require such a bolometric correction because the CR spectrum is expected to be a power law peaking at GeV energies. Then, $L_{\rm CR}|_{\rm SNR}/L_{\rm CR}|_{\rm ISN}\sim \ln(E_{\rm max}/E_{\rm min})M_{\rm ej}\mathcal{R}_{\rm SN}/(M_{\rm CSM}\mathcal{R}_{\rm ISN})$ if we assume SNRs and ISNe have similar values of $\epsilon_{\rm CR}$, $V$, and $\mathcal{R}_{\rm SN}/\mathcal{R}_{\rm ISN}\sim10$. Then, for $M_{\rm ej}\sim10\,M_\odot$ and $M_{\rm CSM}\sim1\,M_\odot$, $L_{\rm CR}|_{\rm SNR}/L_{\rm CR}|_{\rm ISN}\sim 10^3$. Thus, the simple estimate is comparable to the observed ratio, so ISNe warrant more detailed investigation.

\section{\label{sec:SNe}Interacting supernovae}

In this section we detail the shock evolution model, the parameters used to describe Ib, Ic, Ibn, Icn, and IIn ISNe, and the effective temperature of the SN emission, which is important for determining the ionization state of the CSM being injected into the shock.

\subsection{\label{subsec:shock}Shock evolution}

The post-SN ejecta, and the resulting shock, are time-dependent quantities. This results in a time-dependent CR acceleration picture for each supernova. To that effect, we build on the model from Ref.~\cite{Marcowith:2018ifh}, which constructs a self-consistent picture of the evolution of the shock from ISNe. The radius and velocity evolution of the shock can be expressed as
\begin{equation}\label{shockradius}
    R_{\rm sh}=R_0\left(\frac{t}{t_0}\right)^m,
\end{equation}
\begin{equation}\label{shockvelocity}
    V_{\rm sh}=\frac{R_0m}{t_0}\left(\frac{t}{t_0}\right)^{m-1},
\end{equation}
where $R_0$ and $R_0m/t_0=V_0$ is the initial radius and velocity at the initial time, $t_0$, determined by the temporal index $m=(n-3)/(n-s)$, where $n$ is the (negative) power law index of the SN ejecta and $s$ is the (negative) power law index of the CSM density. The CSM density is given by
\begin{equation}\label{csmdensity}
    \rho_{\rm CSM}=D'\left(\frac{R_{\rm sh}}{R_0}\right)^{-s}\times10^{-14}\,{\rm g~cm^{-3}},
\end{equation}
where we take \cite{Maeda:2022xlu}
\begin{equation}\label{dprime}
    D'=10\frac{\dot{M}}{0.05\,{\rm M_{\odot}yr^{-1}}}\frac{100\,{\rm km~s^{-1}}}{V_w}.
\end{equation}
Here, $\dot{M}$ is the pre-supernova mass-loss rate and $V_w$ is the pre-supernova mass-loss (or wind) velocity.

The mass swept up by the shock, $M_{\rm swept}$, is
\begin{align}\label{dmswept}
    \nonumber dM_{\rm swept}&=4\pi\rho_{\rm CSM}R_{\rm sh}^2 dR_{\rm sh},\\
    &=4\pi CR_{\rm sh}^{-s}R_{\rm sh}^2 V_{\rm sh} dt,
\end{align}
where $C=D'R_0^s\times10^{-14}$ is the normalization of the CSM density. For our flux calculation in Sec.~\ref{sec:fluxcomp}, we integrate over all time-dependent quantities from $t_{\rm crit}$, or the initial time that CRs can escape the environment, to $t_f$, the final time. $t_{\rm crit}$ is the time when the characteristic energy of destruction processes is equal to the characteristic energy of escape processes, and is described in more detail in the next section. For our final time, we choose $300\,{\rm days}$ as a fiducial parameter, which corresponds to the outer edge of the extended CSM at $R_{\rm out}\approx4\times10^{16}\,{\rm cm}$. 

The mass swept up depends on the pre-supernova mass loss history and the supernova type, but we check that the total mass swept up for this duration does not exceed the typical SN ejecta mass inferred from observations. For the SN parameters in the next subsection, this $M_{\rm swept}$ ($\sim0.6\,M_\odot$ for IIn SNe, $\sim0.1\,M_\odot$ for Ibn/Icn, and $\sim0.01\,M_\odot$ for Ib/Ic SNe) is much less than typical values of $M_{\rm ej}\sim10\,M_\odot$ \cite{Salmaso:2024jry,Ransome:2024cza} and the assumed shock velocity evolution remains valid over the time we consider here.

The maximum energy that cosmic-rays can reach is related to the magnetic field of the wind, the region where CRs can be accelerated. Assuming the magnetic field energy density is proportional to the kinetic energy density near the stellar surface ($B_w^2/(8\pi)=\rho_w V_w^2/2$, where $\rho_w\sim\dot{M}/(4\pi R_0^2V_w)$), the magnetic field in this region is given by \cite{Marcowith:2018ifh}:
\begin{equation}\label{bfield}
    B_w=\overline{\omega}\left(\frac{\dot{M}V_w}{R_0^2}\right)^{1/2}\left(\frac{t}{t_0}\right)^{-ms/2},
\end{equation}
where $\overline{\omega}=1$ is the ratio of the magnetic field in the wind at $t_0$ to the magnetic field at the surface of the star. The evolution is assumed to be proportional to the density, so $B_w\propto\rho^{1/2}\propto R_{\rm sh}^{-s/2}\propto t^{-ms/2}$. 

This upstream magnetic field can be amplified by various instabilities related to the streaming of cosmic-rays. Ref.~\cite{Marcowith:2018ifh} showed that one of the fastest growing instabilities in ISNe is the non-resonant streaming (NRS), or Bell, instability \cite{Bell04}. In the NRS instability, cosmic-rays generate a current as they escape upstream. This current can grow instabilities in the magnetic field at wavelengths smaller than the Larmor radius. These instabilities grow up to a cutoff wavenumber, at which it is `saturated.' The amplified, saturated magnetic field is proportional to the energy density and the shock velocity, and can be expressed as $B_{\rm sat,NRS}\approx(12\pi\xi_{\rm CR}\rho_{\rm CSM}V_{\rm sh}^3/(\phi c))^{1/2}$ \cite{Pelletier:2006ik,Marcowith:2018ifh}, where $\xi_{\rm CR}$ is the fraction of the shock ram pressure imparted to the nuclei and is proportional to $1/V_{\rm sh}$, which gives its time dependence: $\xi_{\rm CR}=\xi_{\rm CR,0}(t/t_0)^{1-m}$. We take a value of $\xi_{\rm CR,0}=0.05$. $\phi$ is a dimensionless factor set by the minimum and maximum energy of the CR distribution $\phi=\ln(E_{\rm max}/(m_pc^2))$ (in our case, $\phi\approx18$ for a maximum energy of $\sim5\times10^{16}\,{\rm eV}$).

The magnetic field can be amplified up to this saturation magnetic field. This amplification can be labeled, $\mathcal{A}$, given by the ratio of $B_{\rm sat,NRS}/B_w\equiv\mathcal{A}$:
\begin{equation}
\begin{aligned}\label{ampfactor}
    \mathcal{A}&=\max\left(1,\frac{B_{\rm sat,NRS}}{B_w}\right),\\
    &=\max\left(1,\right.\\
    &~~~\left.3.45\times10^{-15}\left(\frac{\xi_{\rm CR,0}}{\phi c}\right)^{1/2}\frac{R_0V_0^{3/2}}{\overline{\omega}V_w}\left(\frac{t}{t_0}\right)^{m-1}\right).
\end{aligned}
\end{equation}
Here, $3.45\times10^{-15}$ comes from the $\rho_{\rm CSM}$ prefactor and the factor of $12\pi$. For $n=7$, $s=2.8$, and $m\approx0.95$, $\mathcal{A}$ is approximately constant in time but for $s=2.4$, $m\approx0.87$, so $\mathcal{A}\propto t^{-0.13}$. For Ib/Ic/Ibn/Icn supernovae with our considered values, $B_{\rm sat,NRS}<B_w$, so $\mathcal{A}=1$ and magnetic field amplification does not occur. For IIn supernovae, $3\lesssim\mathcal{A}\lesssim5$ over the time considered here. Additionally, from the expression $(\mathcal{A}B)^2/8\pi=\epsilon_B\rho_{\rm CSM}V_{\rm sh}^2/2$, the magnetic field efficiency has a value of $\epsilon_B\approx10^{-3}$ for the IIn case with amplified magnetic field. This is similar to the value of $\epsilon_B\lesssim10^{-3}$ from the hybrid simulations of ion acceleration in amplified magnetic fields \cite{Caprioli:2014tva} and may be consistent with radio observations of IIP supernovae \cite{Chevalier:2005aa}.

\subsection{\label{subsec:sntypes}Supernova types}

There are several subtypes of SNe that fall under the ISNe umbrella that arise from periods of pre-supernova mass loss. Here, we describe the model parameters required from each type to predict the resulting CR flux. These are fiducial parameter choices based on inferred properties from multiwavelength observations of ISN lightcurves. In nature, these quantities can vary greatly between individual SNe. In this work, we consider the CR contribution from Ib, Ic, Ibn, Icn, and IIn supernovae. We do not consider supernovae that interact with a confined CSM (\cite{Forster:2018mib,Bruch:2022aqd}, see Sec.~\ref{sec:disc} for more discussion). We choose $R_0=5\times10^{14}\,{\rm cm}$, $t_0=3\,{\rm days}$, $n=7$ \cite{Maeda:2022xlu,Matsuoka:2025lnh}, and $s=2.8$ \cite{Dwarkadas2012,Maeda:2022xlu,Jacobson-Galan:2025bss}. These parameters can vary for individual SN events, so we take these to be representative, fiducial parameters. For these, $m\approx0.95$. The parameters we change for different SN types that are required to estimate the CR contribution are the pre-supernova mass-loss rate, $\dot{M}$, the pre-supernova mass-loss velocity, $V_w$, the nuclear composition, and the fraction of the total supernova rate for that SN type, $f_{\rm SN}$.

First, we consider Ib and Ic SNe - stripped-envelope supernova that are rich in He and C, respectively. Because these do not have narrow-lines, there is likely a less massive CSM surrounding these objects. Some analyses put $\dot{M}$ and $V_w$ at $\gtrsim10^{-3}\,{\rm M_{\odot}~yr^{-1}}$ and $\sim1000\,{\rm km~s^{-1}}$, respectively \cite{Nagy:2025rgi,Dwarkadas:2025gld}. However, the lightcurve is dominated by Ni-decay powered instead of the luminosity of the CSM interaction. With $\dot{M}\sim10^{-3}\,{\rm M_{\odot}~yr^{-1}}$, the luminosity from the CSM-ejecta interaction (described below) is much less than the luminosity from Ni-decay powered lightcurve. We optimistically raise $\dot{M}$ to $\sim10^{-2}\,{\rm M_{\odot}~yr^{-1}}$, because the lightcurve component from the CSM interaction is still less than the component from Ni decay, even after this increase. We model these two luminosity components below.

We can estimate the luminosity from the CSM-ejecta interaction as \cite{Moriya:2013hka}:
\begin{equation}\label{lum}
    L_{\rm CSM}=2\pi\epsilon_{\gamma}R_{\rm sh}^2\rho_{\rm CSM}V_{\rm sh}^3,
\end{equation}
where we assume $\epsilon_{\gamma}$ follows the mean efficiency curve of Figure 4 from Ref.~\cite{Tsuna:2019srj}. The Ni-decay powered luminosity from the SN is estimated by (see, e.g., \cite{Rybicki79}):
\begin{equation}\label{nickellum}
    L_{\rm SN}=e^{-\frac{t^2+2t_0t}{2t_{\rm ch}^2}}\left(\int_0^te^{\frac{t'^2+2t_0t'}{2t_{\rm ch}^2}}\dot{Q}(t')\frac{t'+t_0}{t_{\rm ch}^2}dt'+\frac{E_{\rm in}R_{\rm in}}{t_{\rm ch}^2v_{\rm ej}}\right),
\end{equation}
where $t_{\rm ch}=(2\kappa M_{\rm ej}/\beta cv_{\rm ej})^{1/2}$ \cite{Arnett82}, $\kappa=0.1\,{\rm cm^2~g^{-1}}$ is the opacity of the supernova ejecta \cite{Sutherland84}, $\beta=8$ is a geometric factor, $v_{\rm ej}\sim10,000\,{\rm km~s^{-1}}$ is the velocity of the ejecta, and $M_{\rm ej}\sim3\,{\rm M_{\odot}}$ is the mass ejected. $\dot{Q}=f_{\rm therm}L_{\rm decay}$ where $f_{\rm therm}=1-e^{-\tau_g}$, $\tau_g={3\kappa_gM_{\rm ej}/4\pi R_{\rm ej}^2}$ with $\kappa_g=0.03\,{\rm cm^{2}~g^{-1}}$ \cite{Sutherland84} and $R_{\rm ej}=V_{\rm ej}t$. $L_{\rm decay}$ is given by
\begin{align}\label{ldecay}
    \nonumber L_{\rm decay}&=(6.45\times10^{43}e^{-t/\tau_{\rm Ni}}+1.45\times10^{43}e^{-t/\tau_{\rm Co}})\\
    &\times\left(\frac{M_{\rm Ni}}{1\,{\rm M_{\odot}}}\right)\,{\rm erg~s^{-1}}.
\end{align}
We choose a Nickel mass of $0.2\,{\rm M_{\odot}}$ for Ib/Ic supernovae \cite{Drout11,Anderson:2019yri}. Finally, we choose $E_{\rm in}=M_{\rm ej}v_{\rm ej}^2/2=3\times10^{51}\,{\rm erg}$ and $R_{\rm in}=1\,{\rm R_{\odot}}$. Here, as stated in the previous paragraph, we choose the $\dot{M}$ parameter such that the resulting $L_{\rm SN}>L_{\rm CSM}$. A value of $\dot{M}\sim10^{-2}\,{\rm M_{\odot}~yr^{-1}}$ and $V_w\sim1000\,{\rm km~s^{-1}}$ still satisfies this condition (and is similar to the values reported in Ref.~\cite{Baer-Way:2025amy}).

We assume the nuclear composition of Ib and Ic supernovae is dominantly He and C, respectively. We choose $f_{\rm SN}=11\%$ and 7\% \cite{Shivvers:2016bnc} (also see \cite{2011MNRAS.412.1522S,Toshikage:2024rzp}) of the total local supernova rate for Ib and Ic, respectively.

Next we consider Ibn and Icn supernova. These, like their Ib and Ic counterparts, are also dominantly composed of He and C nuclei. However, narrow line emission comes from a larger component of pre-shocked CSM. This implies a larger mass-loss rate than Ib and Ic supernovae and also that the light curve is powered more by the CSM interaction than nickel decay, i.e., $L_{\rm SN}<L_{\rm CSM}$. We assume these SNe have $\dot{M}\sim10^{-1}\,{\rm M_{\odot}~yr^{-1}}$ and $V_w\sim1000\,{\rm km~s^{-1}}$ \cite{Maeda:2022xlu,Pellegrino:2024zsw,Farias:2025awn}, however with these parameters and $\sim0.2\,{\rm M_{\odot}}$ of Ni ejected, $L_{\rm CSM}>L_{\rm SN}$ initially but becomes comparable at $\sim10\,{\rm days}$. Lastly, we assume that these supernovae combined make up $f_{\rm SN}\sim1\%$ \cite{Maeda:2022xlu,Toshikage:2024rzp} of the total local supernova rate.

Lastly, we consider IIn supernovae. These supernovae also have a prominent CSM as evidenced by narrow line emission, but still show H emission. Thus, these objects are not largely stripped and we assume their elemental abundances match solar composition \cite{Grevesse96}; we take H, He, CNO elements, and Fe as representative nuclei of the solar composition. We choose values of $\dot{M}\sim5\times10^{-2}\,{\rm M_{\odot}~yr^{-1}}$ and $V_w\sim100\,{\rm km~s^{-1}}$ (see Refs.~\cite{Kiewe12,Fransson:2013qya,Kumar:2019tlm,Tartaglia:2019rvg,Moriya:2020gxa,Ransome:2024cza}). Finally, for IIn supernovae, we choose $f_{\rm SN}=5\%$ \cite{Shivvers:2016bnc,Cold:2023cps}. A summary of the physical parameters relevant for each supernova type is shown in Table~\ref{tab:snparams}.

\renewcommand{\arraystretch}{1.5}
\begin{table}[b]
\caption{\label{tab:snparams}Summary of SN parameters including SN type, pre-supernova mass-loss rate ($\dot{M}$), pre-supernova mass-loss wind velocity ($V_w$), fraction of the total local supernova rate ($f_{\rm SN}$), and the nuclear composition (`Comp'). Here, `Solar' composition in our model means the solar abundances of H, He, CNO species, and Fe. Because we assume the same $R_0=5\times10^{14}\,{\rm cm}$, $t_f=300\,{\rm days}$, $n=7$, and $s=2.8$, we get the same outer edge of the CSM ($R_{\rm out}=R_{\rm sh}(300\,{\rm days})\approx4\times10^{16}\,{\rm cm}$) for all types.}
\begin{ruledtabular}
\begin{tabular}{ccccc}
SN Type&$\dot{M}$&$V_w$&$f_{\rm SN}$&Comp\\
&[${\rm M_{\odot}~yr^{-1}}$]&$[{\rm km~s^{-1}}]
$&$[\%]$&\\
\colrule
Ib&$10^{-2}$&$1000$&11&He\\
Ic&$10^{-2}$&$1000$&7&C\\
Ibn&$10^{-1}$&$1000$&0.5&He\\
Icn&$10^{-1}$&$1000$&0.5&C\\
IIn&$5\times10^{-2}$&$100$&5&Solar\\
\end{tabular}
\end{ruledtabular}
\end{table}

\subsection{\label{subsec:temperature}Temperature evolution}

The photon temperature in the vicinity of the acceleration region can play a vital role in the composition of the escaping cosmic-rays because of both photodisintegration and photoionization. Higher photon temperatures result in increased photodisintegration (or a lower achievable maximum energy) for the accelerated nuclei. In some of the cases (like the IIn He nuclei in Fig.~\ref{fig:emaxIInHe} of Sec.~\ref{sec:acceleration}), photodisintegration can limit the maximum energy nuclei can achieve. The temperature also results in more highly photoionized nuclei in the CSM. The ionization state is an important quantity that determines the escape cosmic-ray composition because of the preferential injection effect of singly-ionized nuclei from Ref.~\cite{Caprioli:2017oun}. As a result of their ab initio simulations, heavy nuclei can undergo a large increase in abundance post-injection, depending on their ionization state. For Ib/Ic/Ibn/Icn supernovae, we assume a one-component composition so the following discussion is not important, but is extremely important for the IIn supernova case. In the IIn case, because we assume a four-component, solar-like composition, the observed cosmic-ray composition is influenced by the delicate balance of photodisintegration and photoionization.

Because the photon distribution around the shock is not completely thermal, the temperature evolution of ISNe is not well understood and can vary significantly and/or be underestimated in the $\mathcal{O}$(10 days) that follow the core-collapse \cite{Kumar:2019tlm,Tartaglia:2019rvg,Moriya:2020gxa,Hiramatsu:2024mjz}. To this effect, we choose two different photon temperatures to properly account for the effect of uncertainty of temperature evolution, assuming that it is related to the luminosity. We first demand that the energy density is proportional to $L_{\rm CSM}+L_{\rm SN}$, such that $L_{\rm CSM}+L_{\rm SN}=4\pi R_{\rm sh}^2(\epsilon\sigma_{\rm SB}T^4)$, where $\epsilon$ is the emissivity for the Planck distribution (whose distribution, $dn/d\varepsilon$, is stated explicitly in Sec.~\ref{subsec:maxenergies}). For our first model, $T1$, we assume $\epsilon=1$, resulting in a black body temperature dependent on CSM parameters. The temperature around $t_{\rm crit}$, or the critical time when nuclei can begin escaping the shock, in the IIn case for H/He/CNO/Fe nuclei (at $\sim 7/19/23/28\,{\rm days}$, described in Sec.~\ref{sec:acceleration}) is $\sim 14300/6600/5500/4800\,{\rm K}$. In IIn supernovae, this temperature comes predominantly from the CSM interaction \cite{Smith:2006dk}, but the nickel contribution is non-negligible in some cases, so is included in the luminosity calculation.

However, the temperature in the vicinity of the shock may remain high due to CSM interaction. X-ray photons produced in the shock may not be fully reprocessed and raise the effective temperature that nuclei are exposed to (we discuss this assumption further in Sec.~\ref{sec:disc}). Further, the photon spectrum may not be well described by a black body, which introduces some uncertainty. Thus, we also introduce a second temperature model, $T2$, where we choose a constant temperature of 15000 K. This is reasonable for some IIn supernovae at $\sim10\,{\rm days}$ \cite{Fassia:2000ee,Kumar:2019tlm,Tartaglia:2019rvg}. To maintain consistency of the relationship between $L_{\rm CSM}$ and temperature, the emissivity, $\epsilon$, is reduced in a time-dependent way. That is, $\epsilon=L_{\rm CSM}/(4\pi R_{\rm sh}^2\sigma_{\rm SB}(15000\,{\rm K})^4)$. Note, in this case we assume the temperature is only proportional to $L_{\rm CSM}$ (not $L_{\rm CSM}+L_{\rm SN}$) because, in this estimation, the thermalization is assumed to be incomplete. Although this temperature is not reasonable after several 10s of days, we are primarily concerned with the temperature at the critical escape time (between $\sim7\,{\rm days}$ for H and $\sim30\,{\rm days}$ for Fe). Assuming temperature evolution with these two methods allows us to understand the range in expected effects of photodisintegration and photoionization. The temperatures, other physical parameters at $t_{\rm crit}$, and other models are described in Table~\ref{tab:cloudyinputs} and will be discussed more in Sec.~\ref{sec:acceleration}.

\begin{table}[b]
\caption{\label{tab:cloudyinputs} \texttt{Cloudy} input parameters required to run photoionization calculations for the four IIn models considered in this work. `Temp' is the the temperature, $L_{\rm CSM}$ is the luminosity from the CSM interaction, $R_{\rm sh}$ is the radius of the shock, and `$n_H$' is the H density at the shock radius. All values are calculated at corresponding escape times, $t_{\rm crit}.$}
\begin{ruledtabular}
\begin{tabular}{cccccc}
Model&A/Z&Temp&log($L_{\rm CSM}$)&log($R_{\rm sh}$)&log($n_H$)\\
&&[K]&[erg ${\rm s^{-1}}$]&[cm]&[${\rm cm^{-3}}$]\\
\colrule
$T1$&1/1&14300&43.4&15.0&9.8\\
&4/2&6600&42.8&15.5&8.7\\
&14/7&5500&42.7&15.5&8.4\\
&56/26&4800&42.6&15.6&8.2\\
\hline
$T2$&1/1&15000&43.4&15.0&9.8\\
&4/2&15000&42.7&15.5&8.5\\
&14/7&15000&42.6&15.6&8.3\\
&56/26&15000&42.5&15.7&8.1\\
\hline
$s=2.4$&1/1&15000&42.8&15.4&9.1\\
&4/2&15000&42.4&15.7&8.4\\
&14/7&15000&42.4&15.8&8.2\\
&56/26&15000&42.3&15.8&8.1\\
\hline
Amp off&1/1&15000&42.8&15.5&8.6\\
&4/2&15000&42.7&15.5&8.4\\
&14/7&15000&42.5&15.7&8.1\\
&56/26&15000&42.4&15.8&7.8\\
\end{tabular}
\end{ruledtabular}
\end{table}

\section{\label{sec:acceleration}Particle injection and acceleration}

For efficient CR particle acceleration to occur, the shock must be collisionless \cite{Blandford87,Levinson:2007rj,Murase:2010cu}. However, if the CSM density and optical depth is too high, the shock is radiation-mediated. We evaluate the optical depth of the CSM in order to test this condition for the supernova types and different CSM power-indices, $s$, we consider (types Ib/Ibn/Ic/Icn/IIn and $s=2.4,~2.8$). The optical depth of the un-shocked CSM can be expressed as:
\begin{equation}\label{radmediatedopticaldepth}
    \tau_{\rm CSM}=\int_{R_{\rm sh}}^{R_{\rm out}}\kappa_s\rho_{\rm CSM}(R)dR=\int_t^{t_f}\kappa_s\rho_{\rm CSM}(t)V_{\rm sh}(t)dt
\end{equation}
where $\kappa_s=0.34\,{\rm cm^2/g}$ is the opacity for Compton scattering \cite{Rybicki79}, assuming a fully ionized solar composition. For IIn supernovae (both $s=2.4$ and 2.8), $\tau_{\rm CSM}$ becomes low (less than $c/V_{\rm sh}$) around $3\,{\rm days}$. This is around $t_0$ and before $t_{\rm crit}$ for all cases. For all other supernova types, $\tau_{\rm CSM}<c/V_{\rm sh}$ always, so efficient CR acceleration can occur in all the models we consider.

\subsection{\label{subsec:abundances}CR injection of nuclei}

The ab initio simulations of Ref.~\cite{Caprioli:2017oun,Caprioli:2025lor} reveal that the abundance of nuclei can be increased by a large factor related to their mass and ionization number. It is found that partially ionized nuclei are preferentially injected during DSA by a factor of $(A_i/Q_i)^2$ where $A_i$ and $Q_i$ are the mass number and ionization state, respectively, of nucleus $i$. However, we also assume that injection occurs at non-relativistic energies ($\sim A_im_pc^2$), so the energy density is further multiplied by another factor of $(A_i/Q_i)^{1/2}$ \cite{Caprioli:2017oun,Kimura:2018ggg}.

To determine the amount of abundance enhancement from this effect, we take the ionization state, $Q_i$, of each nuclei in the IIn case at $t_{\rm crit}$. In order to estimate this, we use the photoionization simulation code \texttt{Cloudy} \cite{cloudy25}. The minimum set of required input parameters to run the code for our application are the temperature (given by model $T1$ or $T2$), the luminosity of the ionizing source (given by $L_{\rm CSM}(t_{\rm crit})$), the radius (given by $R_{\rm sh}(t_{\rm crit})$), and the H density (given by $\rho_{\rm CSM}(t_{\rm crit})/m_p$). These input parameters are given in Table~\ref{tab:cloudyinputs} for each model. 

The resulting fraction of the nuclei in the neutral, singly-, and doubly-ionized state and the modified abundances are given in Table~\ref{tab:cloudyfractions}. After the results from the \texttt{Cloudy} calculations, the modified fraction can be expressed as $f_i=Y_{i,\odot}\left[f_{i,\rm inj,1}(A_i/1)^{5/2}+f_{i,\rm inj,2}(A_i/2)^{5/2}+...\right]/{Norm}$, where $Y_{i,\odot}$ is the solar abundance of species $i$ (for the CNO species, we take the sum of the C, N, and O solar abundances). \textit{Norm} is determined such that $\Sigma_if_i=1$. With this set of parameters, from Table~\ref{tab:cloudyinputs}, H and Fe are easily ionized at $\sim4000-5000\,{\rm K}$, but CNO requires a higher temperature to ionize. At even higher temperatures, the H and CNO nuclei remain mostly singly-ionized while Fe becomes mostly doubly-ionized because of its lower ionization potential. However, He has a much higher ionization potential, so even temperatures of $15000\,{\rm K}$ are not sufficient to totally singly-ionize He. When He is largely ionized, this leads to a larger contribution in the modified fraction $f_i$. This photoionization effect is the cause of the large increases in abundance for the CNO nuclei and Fe.

Also in Table~\ref{tab:cloudyfractions} we show the results for two additional models: $s=2.4$ and `Amp off.' In $s=2.4$, we retain the temperature from the $T2$ model but decrease the CSM density power-law index from 2.8 to 2.4 to account for some of the variation in the density profile index of ISNe (see, e.g., \cite{Maeda:2022xlu,Chiba:2024ksb,Ransome:2024cza}). This decrease results in a later $t_{\rm crit}$ compared to $T2$. In `Amp off,' we again consider a temperature of $15000\,{\rm K}$, but no longer consider the NRS instability mechanism that amplifies the magnetic field. This case also results in a slightly later $t_{\rm crit}$ compared to $T2$. The fraction of singly-ionized He decreases from the `Amp off' model to the others because the former model has a higher ionization state. For a fixed temperature, the ionization state is $\xi\propto L_{\rm CSM}/(n_HR_{\rm sh}^2)$, which is highest in `Amp off,' followed by $T2$, followed by $s=2.4$.

\begin{table}[b]
\caption{\label{tab:cloudyfractions}Fraction of neutral atoms ($f_{i,\rm neut}$), single ($f_{i,\rm inj,1}$) and double ($f_{i,\rm inj,2}$) ionization state fractions, and modified abundance fraction ($f_i$) after the preferential injection effect. These are calculated using the \texttt{Cloudy} input parameters from Table~\ref{tab:cloudyinputs} for each of our four IIn models. The He nuclei in these models are only partially singly ionized; the remaining fraction of He in each model is neutral so are not efficiently accelerated. $f_i$ is normalized in each case such that $\Sigma_if_i=1$.}
\begin{ruledtabular}
\begin{tabular}{cccccc}
Model&&H&He&CNO&Fe\\
\colrule
$T1$&$f_{i,\rm neut}$&0.0&1.0&0.6&0.0\\
&$f_{i,\rm inj,1}$&1.0&0.0&0.4&1.0\\
&$f_{i,\rm inj,2}$&0.0&0.0&0.0&0.0\\
&$f_i$&0.52&0.00&0.18&0.30\\
\hline
$T2$&$f_{i,\rm neut}$&0.0&0.3&0.0&0.0\\
&$f_{i,\rm inj,1}$&1.0&0.7&1.0&0.0\\
&$f_{i,\rm inj,2}$&0.0&0.0&0.0&1.0\\
&$f_i$&0.24&0.53&0.20&0.02\\
\hline
$s=2.4$&$f_{i,\rm neut}$&0.0&0.9&0.0&0.0\\
&$f_{i,\rm inj,1}$&1.0&0.1&1.0&0.2\\
&$f_{i,\rm inj,2}$&0.0&0.0&0.0&0.8\\
&$f_i$&0.42&0.13&0.36&0.08\\
\hline
Amp off&$f_{i,\rm neut}$&0.0&0.2&0.0&0.0\\
&$f_{i,\rm inj,1}$&1.0&0.8&1.0&0.0\\
&$f_{i,\rm inj,2}$&0.0&0.0&0.0&1.0\\
&$f_i$&0.22&0.57&0.19&0.02\\
\end{tabular}
\end{ruledtabular}
\end{table}

\subsection{\label{subsec:maxenergies}Maximum energies}

Once these supernova shocks become collisionless, the maximum energy of cosmic rays is determined by the competition of acceleration and destruction processes. We build on the model expressed in Ref.~\cite{Marcowith:2018ifh} to consider nuclei in addition to protons and to consider the photodisintegration process. Altogether, we consider acceleration, spallation of nuclei and $pp$ interactions in the case of protons, adiabatic expansion, photodisintegration, geometric escape, and the effect of magnetic field amplification due to escaping CRs. In this section, we use the notation $F_x\equiv F/10^x$ in cgs units.

We first discuss the acceleration timescale. In DSA, the diffusion timescale is balanced with the advection timescale. The diffusion timescale is given by $t_{\rm diff}=l_{\rm diff}^2/\kappa_u$, where $l_{\rm diff}$ is the diffusion length and $\kappa_u=\eta r_lc/3$ is the upstream diffusion coefficient and $r_l=E/(Ze\mathcal{A}B_w)$ is the Larmor radius. The advection timescale is written as $t_{\rm adv}=l_{\rm diff}/V_{\rm sh}$. When $t_{\rm adv}=t_{\rm diff}$, we get that $l_{\rm diff}=\kappa_u/V_{\rm sh}$. Plugging this back in gives us $t_{\rm adv}=\kappa_u/V_{\rm sh}^2$. Finally, the acceleration timescale is proportional to the CR advection timescale, such that $t_{\rm acc}=g(r)t_{\rm adv}$, and can be expressed as
\begin{align}\label{taccel}
    \nonumber t_{i,\rm acc}&=E_i\frac{g\eta c}{3Z_ie\mathcal{A}B_wV_{\rm sh}^2}\\
    \nonumber&=\frac{E_i}{\mathcal{A}}\frac{g\eta cR_0}{3Z_ie\overline{\omega}V_0^2\dot{M}^{1/2}V_w^{1/2}}\left(\frac{t}{t_0}\right)^{2(1-m)+ms/2}\\
    \nonumber&\approx1.5\times10^5\,{\rm s}\frac{g_1\eta_0}{\overline{\omega}_0V_{0,9}^2V_{w,7}^{1/2}}\left(\frac{t}{t_0}\right)^{2(1-m)+ms/2}\\
    \nonumber&\times\left(\frac{E_i}{5\times10^{16}\,{\rm eV}}\right)\left(\frac{R_0}{5\times10^{14}\,{\rm cm}}\right)\\
    &\times\left(\frac{\mathcal{A}}{5}\right)^{-1}\left(\frac{Z_i}{2}\right)^{-1}\left(\frac{\dot{M}}{5\times10^{-2}\,{\rm M_\odot~yr^{-1}}}\right)^{-1/2}
\end{align}
where $g=3r(1+r/r_B)/(r-1)$ depends on the ratio of the downstream and upstream magnetic field strengths $r_B\sim\sqrt{(1+2r^2)/3}$, and the shock compression ratio $r\approx4$ for strong adiabatic shocks in the test particle limit \cite{1988ApJ...333L..65V,BEREZHKO1996367}, $\eta=1$ is a diffusion parameter that depends on the magnetic field configuration, and $E_i$ and $Z_i$ are the energy and the atomic number for species $i$. This timescale is compared to the destruction process timescales below in order to determine maximum CR energies.

Note that we assume nuclei are injected in partially ionized states, but become fully stripped of their electrons once accelerated (i.e., the $Z_i$ refers specifically to the atomic number of species $i$). This assumption is reasonable if the ionization timescale is larger than $t_{\rm acc}$ during injection at nonrelativistic energies, but is faster than acceleration once nuclei reach larger energies. We confirm that this is the case for collisional- and photo-ionization.

The first destruction process we consider is due to nuclear processes. This timescale can be expressed as $t_{i,\rm nuc}=(K_{\rm nuc}\sigma_{\rm nuc}\left<n_H\right>c)^{-1}$. As in Ref.~\cite{Marcowith:2018ifh}, we choose an inelasticity of $K_{\rm nuc}\sim0.5$ for protons but $K_{\rm nuc}\sim1/A$ for nuclei. The average number density, $\left<n_H\right>=4Fn_H=4F\dot{M}/(4\pi V_wR_0^2\mu m_p)$, is weighted by the nucleus's residence time in the up and down stream expressed through the factor $F=(1+r_B/(4r)/(1+r_B/r)$. $\mu$ is the mean molecular weight, so is equal to 1.3, 4, and 12 for IIn (at solar composition), Ib/Ibn, and Ic/Icn supernovae, respectively. For the nuclear destruction cross section, $\sigma_{\rm nuc}$, we similarly assume it is approximately constant at very high energies with a value of $\sim5.61\times10^{-26}\,{\rm cm^2}$ for protons to represent $pp$ collisions \cite{Marcowith:2018ifh}. For nuclei with mass $A$, we assume the spallation cross section $\sigma_{\rm nuc}\sim\sigma_{\rm sp}$ \cite{Letaw83} where
\begin{align}
    \sigma_{\rm sp}&=45A^{0.7}(1+0.016\sin(5.3-2.63\ln A)\,{\rm mb}.
\end{align}
This cross section is the constant approximation at high energies ($\gtrsim1\,{\rm GeV}$), but has energy dependence at lower energies. Further, if $A=4$, this is multiplied by 0.8. If $A=9$, this is also multiplied by an energy dependent factor, but is approximately 1 at high energies. This gives a final nuclear destruction timescale of
\begin{align}\label{tnuc}
    \nonumber t_{i,\rm nuc}&=\frac{\pi V_wR_0^2}{\dot{M}}\frac{\mu m_pA_i}{F\sigma_{\rm nuc}c}\left(\frac{t}{t_0}\right)^{ms}\\
    \nonumber&\approx1.1\times10^4\,{\rm s}\frac{V_{w,7}}{F_0\sigma_{\rm nuc,-25}}\left(\frac{t}{t_0}\right)^{ms}\left(\frac{R_0}{5\times10^{14}\,{\rm cm}}\right)^2\\
    &\times\left(\frac{\mu}{1.3}\right)\left(\frac{A_i}{4}\right)\left(\frac{\dot{M}}{5\times10^{-2}\,{\rm M_\odot~yr^{-1}}}\right)^{-1}.
\end{align}
Solving $t_{\rm acc}=t_{\rm nuc}$ for energy gives the expression
\begin{align}\label{enuc}
    \nonumber E_{i,\rm nuc}&=\mathcal{A}\frac{3\pi V_w^{3/2}R_0V_0^2\mu m_pA_iZ_ie\overline{\omega}}{\dot{M}^{1/2}gF\sigma_{\rm nuc}c^2\eta}\left(\frac{t}{t_0}\right)^{2(m-1)+ms/2}\\
    \nonumber &\approx 2.4\times10^{15}\,{\rm eV}\frac{V_{w,7}^{3/2}V_{0,9}^2\overline{\omega}_0}{g_1F_0\sigma_{\rm nuc,-25}\eta_0}\left(\frac{t}{t_0}\right)^{2(m-1)+ms/2}\\
    \nonumber &\times\left(\frac{\mathcal{A}}{5}\right)\left(\frac{R_0}{5\times10^{14}\,{\rm cm}}\right)\left(\frac{\mu}{1.3}\right)\left(\frac{A_i}{4}\right)\left(\frac{Z_i}{2}\right)\\
    &\times\left(\frac{\dot{M}}{5\times10^{-2}\,{\rm M_\odot~yr^{-1}}}\right)^{-1/2}.
\end{align}

\begin{figure*}
\centering
\includegraphics[width=0.49\linewidth]{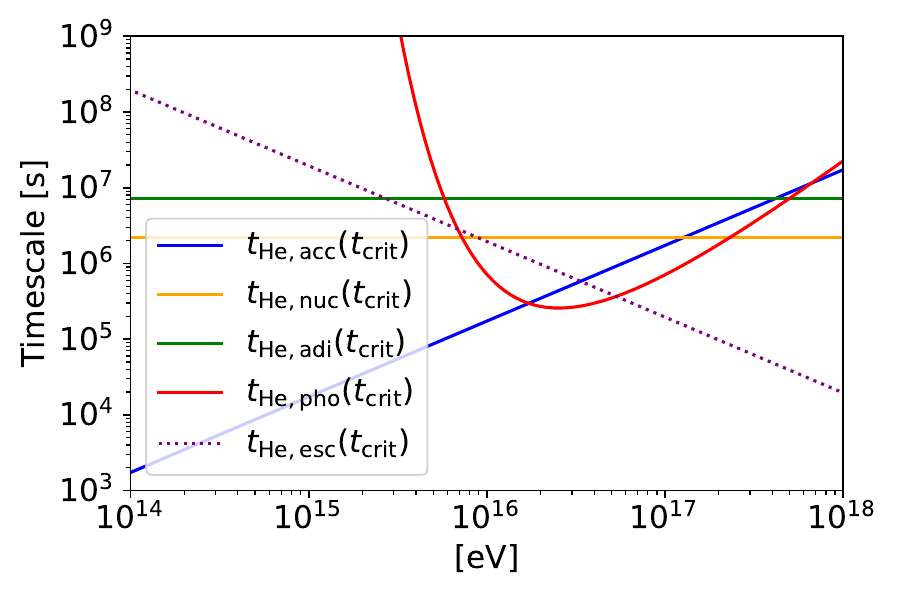}
\includegraphics[width=0.49\linewidth]{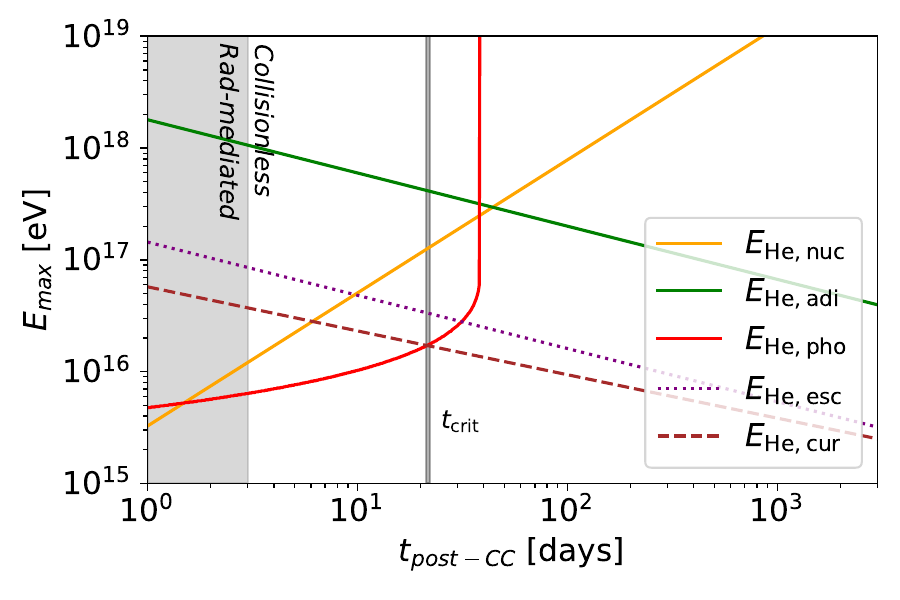}
\caption{These panels show quantities for He nuclei in IIn supernovae in the $T2$ model. \textit{Left panel:} Comparison of the acceleration and energy-limiting timescales at $t_{\rm crit}$. Note that $t_{\rm crit}$ is determined by the magnetic field amplification effect of escaping cosmic rays, but there is no corresponding timescale defined for this. For $t_{\rm pho}$, this is calculated at each time and model, since an analytic expression for $E_{i,\rm pho}$ cannot be written. Photodisintegration is the most restricting destruction process for many cases, prior to the escape time. \textit{Right panel:} Maximum energies achievable for He nuclei. The shaded gray region shows when the radiation-mediated condition is not met (see Eq.~\ref{radmediatedopticaldepth}). After this, around $3\,{\rm days}$, the shock becomes collisionless and CRs can be accelerated. Energy-limiting processes include nuclear processes, adiabatic losses, photodisintegration, geometric escape, and the current induced from escaping nuclei. Here, the dotted lines (for $E_{\rm He,esc}$ and $E_{\rm He,cur}$) represent the processes that allow accelerated nuclei to escape. Here, $s = 2.8$ and magnetic field amplification is considered. In this case, the maximum energy is limited by photodisintegration for the first $\sim22\,{\rm days}$ ($t_{\rm crit}$, the vertical gray line) before escaping upstream. After this point, the maximum energy is limited by the non-resonant streaming instability effect that leads to magnetic field amplification, giving a maximum energy of $\sim10^{16}\,{\rm eV}$. Depending on the composition and input parameters like temperature and power-law index $s$, most cosmic-rays are limited in energy by either nuclear spallation, $pp$ interactions, or photodisintegration.}
\label{fig:emaxIInHe}
\end{figure*}

Next, because of the rapid volume increase, we consider any energy lost through adiabatic expansion. The timescale for adiabatic expansion is given by \cite{1988ApJ...333L..65V,Marcowith:2018ifh}
\begin{align}\label{tadi}
    \nonumber t_{\rm adi}&\approx\frac{3R_0r}{2V_0(r-1)}\left(1+\frac{r_B}{r}\right)\left(\frac{t}{t_0}\right)\\
    \nonumber &\approx 1.8\times10^6\,{\rm s}\frac{1}{V_{0,9}}\left(\frac{R_0}{5\times10^{14}\,{\rm cm}}\right)\left(\frac{t}{t_0}\right)\\
    &\times \left(\frac{r+r_B}{r-1}\right)\left(\frac{1}{2.43}\right).
\end{align}
Solving $t_{\rm acc}=t_{\rm adi}$ for energy gives
\begin{align}\label{eadi}
    \nonumber E_{i,\rm adi}&=\mathcal{A}\frac{3Z_ie\overline{\omega}V_0\dot{M}^{1/2}V_w^{1/2}r_B}{2\eta cr}\left(\frac{t}{t_0}\right)^{2m-1-ms/2}\\
    \nonumber &\approx 7.0\times10^{17}\,{\rm eV}\frac{\overline{\omega}_0V_{0,9}V_{w,7}^{1/2}}{\eta_0}\left(\frac{t}{t_0}\right)^{2m-1-ms/2}\\
    \nonumber &\times \left(\frac{\mathcal{A}}{5}\right)\left(\frac{Z_i}{2}\right)\left(\frac{\dot{M}}{5\times10^{-2}\,{\rm M_\odot~yr^{-1}}}\right)^{1/2}\\
    &\times\left(\frac{r_B}{r}\right)\left(\frac{1}{0.83}\right)
\end{align}

At $\gtrsim{\rm PeV}$ energies, photodisintegration of nuclei may become a relevant process. The photodisintegration timescale is given by
\begin{align}\label{timescale}
    \nonumber t_{i,\rm pho}^{-1}&=\frac{c}{2\gamma_A^2}\int_{\overline{\varepsilon}_{\rm th}}^\infty d\overline{\varepsilon}\ \sigma_{\rm pho}(\overline{\varepsilon})K_{\rm pho}(\overline{\varepsilon})\overline{\varepsilon}\int_{\overline{\varepsilon}/2\gamma_A}^\infty d\varepsilon\ \varepsilon^{-2}\frac{dn}{d\varepsilon}\\
    \nonumber &=\frac{-4\pi c\sigma_{\rm GDR}\Delta\varepsilon_{\rm GDR}\overline{\varepsilon}_{\rm GDR}}{A_i\gamma_A^2(hc)^3}k_BT{\rm ln}(1-e^{-y})\\
    \nonumber &\approx 1.5\times10^{-5}\,{\rm s^{-1}}\sigma_{\rm GDR,-27}\left(\frac{\Delta\varepsilon_{\rm GDR}}{10\,{\rm MeV}}\right)\left(\frac{\overline{\varepsilon}_{\rm GDR}}{30\,{\rm MeV}}\right)\\
    &\times \left(\frac{T}{15000\,{\rm K}}\right)\left(\frac{\ln(1-e^{-y})}{-0.54}\right)\left(\frac{E_i}{5\times10^{16}\,{\rm eV}}\right)^{-2},
\end{align}
for the following assumptions. If the photon spectrum is thermal, $dn/d\varepsilon=8\pi\varepsilon^2/(h^3c^3(e^{\varepsilon/k_BT_{\gamma}}-1))$, $\gamma_A=E_i/(A_im_pc^2)$ is the nucleus Lorentz factor with $A$ as the nucleus mass number, $K_{\rm pho}\sim1/A$ is the nucleus elasticity, $\sigma_{\rm pho}\approx\sigma_{\rm GDR}\delta(\overline{\varepsilon}-\overline{\varepsilon}_{\rm GDR})\Delta\varepsilon_{\rm GDR}$ is the photodisintegration cross section and dominated by the Giant Dipole Resonance (GDR) \cite{KARAKULA1993229}, where $\sigma_{\rm GDR}\sim0.427A^{1.35}\times10^{-27}\,{\rm cm^2}$, $\Delta\varepsilon_{\rm GDR}\sim21.051A^{-0.35}\,{\rm MeV}$, and we assume $\overline{\varepsilon}_{\rm th}\sim1\,{\rm MeV}$. For nuclei with $A<4$, $\overline{\varepsilon}_{\rm GDR}\sim0.925A^{-2.433}\,{\rm MeV}$ and $\sim42.65A^{-0.21}\,{\rm MeV}$ for heavier nuclei [Ekanger \textit{et al} 2026 (in preparation)]. Finally, $y=\overline{\varepsilon}_{\rm GDR}/(2\gamma_Ak_BT)$ for some temperature $T$. We cannot analytically write the maximum energy from the condition $t_{\rm acc}=t_{\rm pho}$, but at each time we find the energy at which these two are equal. The \textit{left panel} of Fig.~\ref{fig:emaxIInHe} shows the acceleration timescale alongside the other destruction processes, like photodisintegration, in addition to the escape timescale. This is shown only at the escape time, $t_{\rm crit}$ (described below), but is carried out for all times and models. Note that $t_{\rm crit}$ is determined by the magnetic field amplification effect due to escaping cosmic rays, but there is no corresponding timescale defined for this.

Finally, we consider the maximum energy of cosmic rays due to two escape processes: geometric losses and the current induced by escaping cosmic rays. These escape processes also determine the time when CRs are first able to escape and are relevant when considering the CR observables.

CRs can escape through geometrical losses due to the expanding volume. The timescale of this process can be written as the diffusion timescale, $t_{\rm diff}$, when the diffusion length becomes a significant fraction of the shock radius, i.e., when $l_{\rm diff}\sim\eta_{\rm esc}R_{\rm sh}$  (or, equivalently, $\kappa_u/V_{\rm sh}=\eta_{\rm esc}R_{\rm sh}$) \cite{BEREZHKO1996367,Marcowith:2018ifh}. This escape timescale can be expressed as
\begin{align}\label{tesc}
    \nonumber t_{i,\rm esc}&=\frac{\eta_{\rm esc}^2R_{\rm sh}^2}{\kappa_u}\\
    \nonumber &=\mathcal{A}\frac{3Z_ie\eta_{\rm esc}^2R_0\overline{\omega}\dot{M}^{1/2}V_w^{1/2}}{E_i\eta c}\left(\frac{t}{t_0}\right)^{2m-ms/2}\\
    \nonumber &\approx1.7\times10^5\,{\rm s}\frac{\eta_{\rm esc,-1}^2\overline{\omega}_0V_{w,7}^{1/2}}{\eta_0}\left(\frac{t}{t_0}\right)^{2m-ms/2}\\
    \nonumber &\times\left(\frac{\mathcal{A}}{5}\right)\left(\frac{Z_i}{2}\right)\left(\frac{R_0}{5\times10^{14}\,{\rm cm}}\right)\\
    &\times\left(\frac{\dot{M}}{5\times10^{-2}\,{\rm M_\odot~yr^{-1}}}\right)^{1/2}\left(\frac{E_i}{5\times10^{16}\,{\rm eV}}\right)^{-1}.
\end{align}
The condition $l_{\rm diff}\sim\eta_{\rm esc}R_{\rm sh}$ directly provides the maximum energy \cite{Marcowith:2018ifh}:
\begin{align}\label{eesc}
    \nonumber E_{i,\rm esc}&=\mathcal{A}\frac{\eta_{\rm esc}Z_ie\overline{\omega}\dot{M}^{1/2}V_w^{1/2}V_0}{\eta c}\left(\frac{t}{t_0}\right)^{2m-1-ms/2}\\
    \nonumber &\approx 5.6\times10^{16}\,{\rm eV}\frac{\eta_{\rm esc,-1}\overline{\omega}_0V_{w,7}^{1/2}V_{0,9}}{\eta_{0}}\left(\frac{t}{t_0}\right)^{2m-1-ms/2}\\
    &\times \left(\frac{\mathcal{A}}{5}\right)\left(\frac{Z_i}{2}\right)\left(\frac{\dot{M}}{5\times10^{-2}\,{\rm M_\odot~yr^{-1}}}\right)^{1/2},
\end{align}
where $\eta_{\rm esc}$ parametrizes the amount of particle loss. This condition is equivalent to setting $t_{\rm esc}=t_{\rm adv}$. If we compare the escape timescale to the acceleration timescale instead of advection, this results in an additional factor of $1/g^{1/2}$ in $E_{i,\rm esc}$.

In this study, we also consider the effect of magnetic field amplification due to the streaming of escaping accelerated nuclei. These escaping nuclei determine the maximum energy of the CR spectrum, $E_{\rm cur}$. Following Ref.~\cite{Schure:2013yga}, we start with the assumption that $N_0E_0=N_{\rm esc}E_{\rm cur}$ for an $E^{-2}$ number spectrum normalized by $N_0/E_0$. Here, $N_0$ represents the number density corresponding to some minimum energy $E_0=m_pc^2$. The CR energy density can be derived from integrating the number spectrum, such that $U_{\rm CR}=N_0E_0\ln(E_{\rm cur}/(m_pc^2))=N_0E_0\phi=\chi\rho_{\rm CSM}V_{\rm sh}^2$, where $\chi$ represents some fraction of the shock's kinetic energy transferred to CRs. The number density of escaping CRs can be related to the current produced by streaming CRs by the relation: $N_{\rm esc}=j/(q\zeta V_{\rm sh})$. Here, the current can be related to the magnetic field amplification factor as $j=\mathcal{N}cV_{\rm sh}\rho_{\rm CSM}^{1/2}/(R_{\rm sh}\pi^{1/2})$ where $\mathcal{N}=\ln\mathcal{A}$ is the number of e-folding times growth of the NRS instability \cite{Schure:2013kya}. The quantity $\zeta V_{\rm sh}$ is the drift velocity of escaping CRs. In order to relate this drift velocity to our SN parameters, we have to introduce the energy flux of escaping CRs, $Q_{\rm esc}$, which have velocity $\zeta V_{\rm sh}$. This energy flux is given by $Q_{\rm esc}=\zeta V_{\rm sh}U_{\rm CR}=\zeta\chi\rho_{\rm CSM}V_{\rm sh}^3$ and is related to the shock ram pressure imparted to CRs, $P_{\rm CR}$, by $Q_{\rm esc}=P_{\rm CR}V_{\rm sh}=\xi_{\rm CR}\rho_{\rm CSM}V_{\rm sh}^3$, such that $\zeta V_{\rm sh}=\xi_{\rm CR}V_{\rm sh}/\chi$.

Finally, we can express the maximum energy achievable with the magnetic field amplification effect, $E_{\rm cur}$. Starting from $E_{\rm cur}=N_0E_0/N_{\rm esc}$, and combining the expression for $N_0E_0$ in terms of the shock's kinetic energy, $N_{\rm esc}$ in terms of the CR current, and $\zeta V_{\rm sh}$ in terms of SN parameters, we can write this maximum energy as \cite{Marcowith:2018ifh}:
\begin{align}
    \nonumber E_{i,\rm cur}&=10^{-15}\frac{Z_ie\xi_{\rm CR,0}R_0V_0^2\dot{M}^{1/2}}{\phi\mathcal{N}cV_w^{1/2}}\left(\frac{t}{t_0}\right)^{2m-1-ms/2}\\
    \nonumber &\approx 9.7\times10^{15}\,{\rm eV}\frac{V_{0,9}^2}{V_{w,7}^{1/2}}\left(\frac{t}{t_0}\right)^{2m-1-ms/2}\\
    \nonumber &\times \left(\frac{Z_i}{2}\right)\left(\frac{\xi_{\rm CR,0}}{0.05}\right)\left(\frac{R_0}{5\times10^{14}\,{\rm cm}}\right)\\
    &\times\left(\frac{\dot{M}}{5\times10^{-2}\,{\rm M_\odot~yr^{-1}}}\right)^{1/2}\left(\frac{\phi}{18}\right)^{-1}\left(\frac{\mathcal{N}}{\ln(5)}\right)^{-1}.
\end{align}
Here, as in $\mathcal{A}$, the $10^{-15}$ prefactor comes from the $\rho_{\rm CSM}$ and the factor of $\pi$.

We are primarily considering the case where the magnetic field is amplified due to CR streaming, in which case $r_B\sim\sqrt{(1+2r^2)/3}$ \cite{Parizot:2006rt}. Without magnetic field amplification, $r_B$ and $\eta$ depend on the magnetic field configuration. If the wind magnetic field is assumed to be parallel to the shock normal, $r_B\sim1$ and the dependence on $\eta$ is shown correctly in the maximum energy expressions. This is because the acceleration timescale is proportional to this diffusion parameter, i.e., $t_{\rm acc}\propto\eta$. We consider this parallel configuration case later in the `Amp off' model. However, if the wind magnetic field is assumed to be perpendicular to the shock normal, $r_B\sim r$. In this case, $t_{\rm acc}\propto1/\eta$ and all expressions are modified accordingly \cite{Marcowith:2018ifh}. Since we assume Bohm diffusion, i.e., $\eta=1$, $t_{\rm acc}$ for the perpendicular case is almost the same as the that for the parallel case.

The maximum energy achievable by a nuclear species in this setup is given by $E_{i,{\rm max}}={\rm Min}(E_{i,{\rm nuc}},~E_{i,{\rm adi}},~E_{i,{\rm esc}},~E_{i,{\rm cur}},~E_{i,{\rm pho}})$. Although nuclei are accelerated within this framework, the time when they can escape and begin propagation in the ISM is given by $t_{\rm crit}=t(E_{i,{\rm max}}={\rm Min}(E_{i,{\rm esc}},~E_{i,{\rm cur}}))$. In other words, the escape time is given by the time when the escape processes become the limiting process in terms of maximum energy. Note that, in the `Amp off' model, we do not consider the case where the NRS instability operates. Therefore, $t_{\rm esc}=t(E_{i,{\rm max}}=E_{i,{\rm esc}})$ for that case.

In the \textit{right panel} of Fig.~\ref{fig:emaxIInHe}, we show the maximum energy that He nuclei in IIn supernovae can reach based on the preceding processes. The gray shaded region up to $\sim3\,{\rm days}$ shows the time when the shock is radiation-mediated, and therefore no particle acceleration occurs. In orange we show the maximum energy from nuclear processes like spallation, in green we show the maximum energy from adiabatic losses, and in red we show the maximum energy where nuclei are limited by photodisintegration. Geometric escape (purple dotted) and magnetic field amplification due to the streaming of escaping nuclei (brown dotted) also limit the maximum energy of accelerated nuclei. For He nuclei in IIn supernovae, $t_{\rm crit}\sim22\,{\rm days}$ (given by the vertical gray line). Here, photodisintegration is the main process that limits maximum achievable energies until current-driven maximum energies are the most limiting after a few 10s of days post-core collapse. This is also true for the $T1$ and $T2$, for all nuclei, but $pp$ interactions are the limiting process for hydrogen in all cases. Similarly, in the `Amp off' model, photodisintegration limits maximum achievable energies, but geometric escape becomes the limiting escape process. The behavior is somewhat different for the $s=2.4$ model and for non-IIn ISNe. In those cases, spallation determines the maximum energy for the first few days, then photodisintegration, then the maximum energy associated with geometric escape, not current-driven, is the most limiting. Overall, the general behavior of these models is that maximum energies are limited by nuclear destruction processes until they can escape. In principle, photo-meson production could be a relevant limiting process. However, for protons and nuclei in all cases, very high Lorentz factors are required to overcome the threshold for the photo-meson resonance and the timescale is always much larger than the photodisintegration and acceleration timescales, so can be safely ignored.

\section{\label{sec:fluxcomp}Cosmic-ray flux and composition}

In this framework, it is feasible for nuclei to be accelerated to super-knee energies but it is equally important to know if these supernova classes can provide the necessary flux of cosmic-rays from observations. Additionally, the predicted composition can be compared to that inferred from observations. Here, we detail how we estimate these quantities and compare them to recent measurements of super-knee cosmic-rays.

We assume that the CR injection spectra are power laws with index 2. The normalization for this injection spectrum and the resulting escape spectrum is proportional to the energy in CRs per event ($\sim\epsilon_{\rm CR}M_{\rm swept}V_{\rm sh}^2/2$) and a bolometric correction factor from integrating the injection spectrum $\int_{E_{p,\rm min}}^{E_{p,\rm max}}E(dN/dE)dE$. For the escape-limited model \cite{Ohira10,Zhang:2017moz}, we assume that only nuclei with energy close to the maximum energy escape. This results in the following (differential) log-normal distribution of the escape spectrum for a single species in a single ISN event:

\begin{figure*}
\centering
\includegraphics[width=0.49\linewidth]{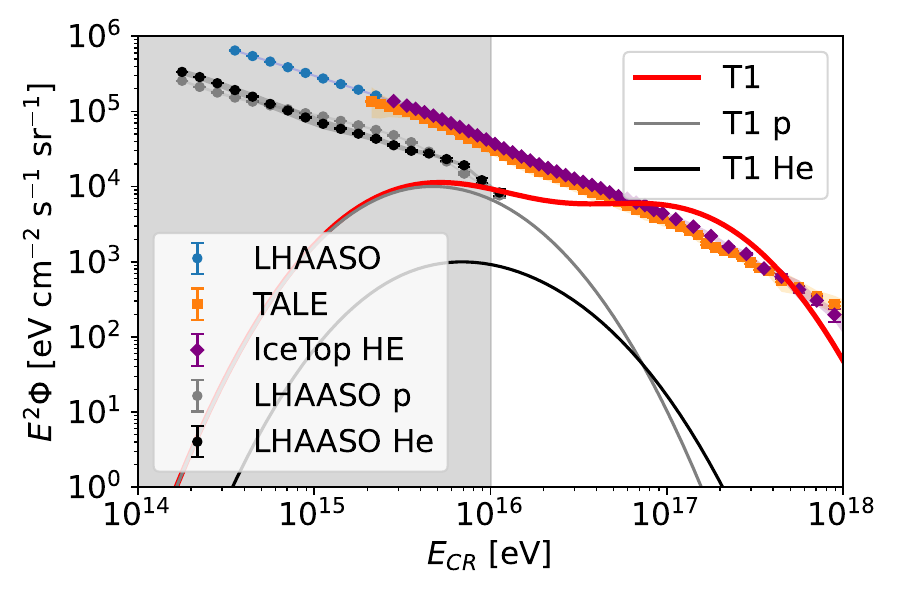}
\includegraphics[width=0.49\linewidth]{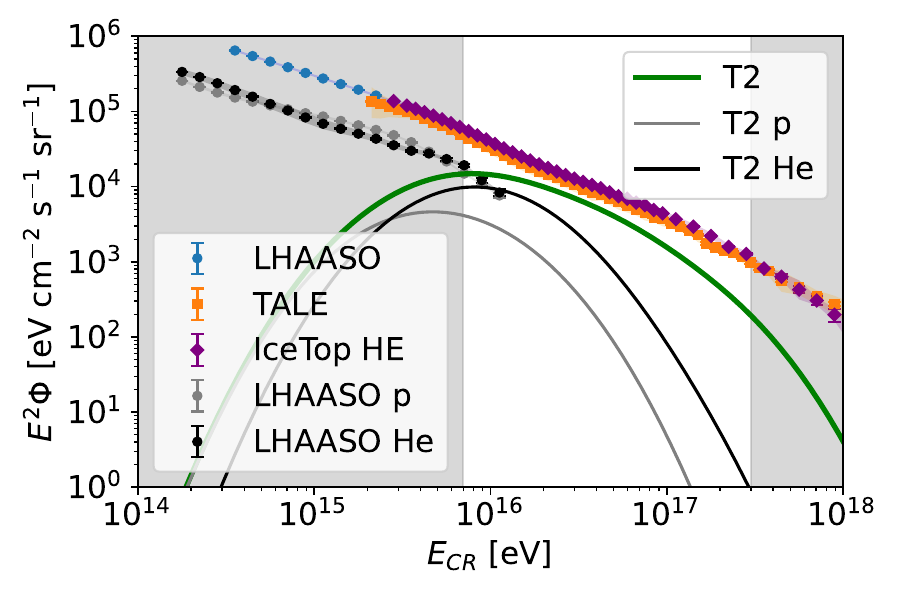}
\includegraphics[width=0.49\linewidth]{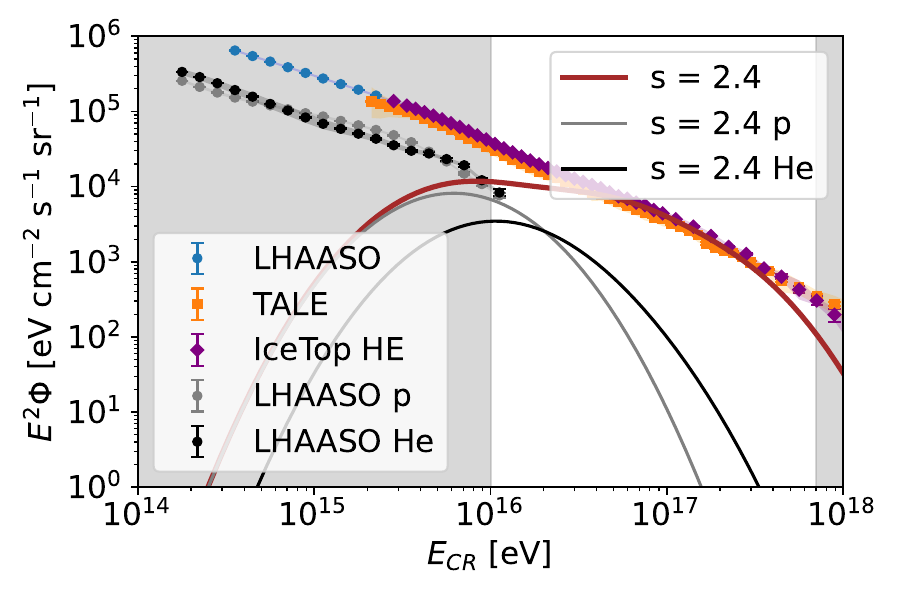}
\includegraphics[width=0.49\linewidth]{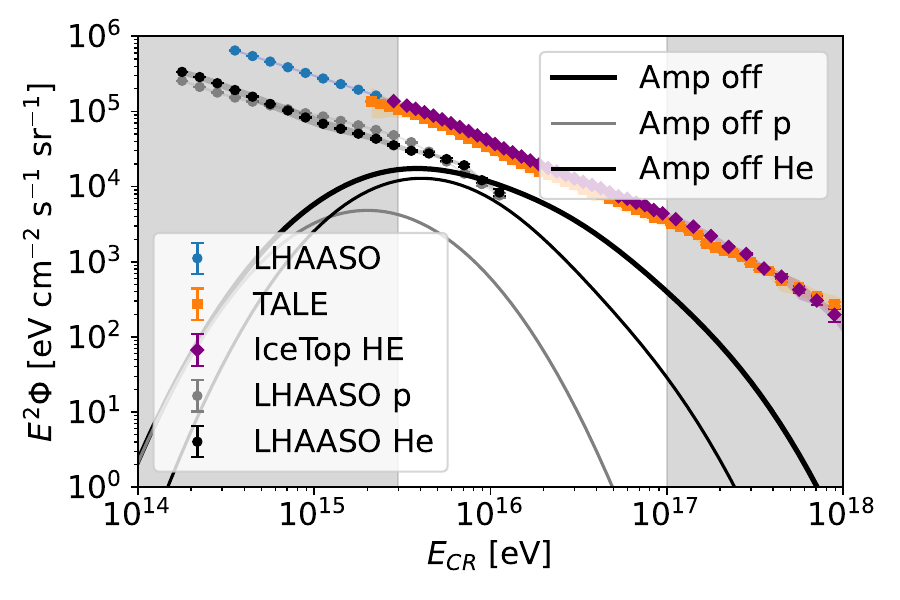}
\caption{Predicted flux from different models compared to observed fluxes from LHAASO (and the LHAASO p- and He-only spectra), TA, and IceTop, shown in the thinner line styles. Statistical errors are shown as vertical bars and systematic errors are shown as a shaded region. We highlight the white regions in each panel where these models (thicker lines) generally provide a dominant fraction of the observed flux. All models are consistent with the recent LHAASO p- and He-only fluxes, but the relative compositions and overall flux are useful in constraining models (see also Appendix~\ref{sec:icetopresults}). \textit{Upper left panel:} shows the predicted flux for the $T1$ model, where $T\sim L^{1/4}$. Although, consistent with LHAASO results, this model overestimates the observed flux at $10^{17}\,{\rm eV}$ because of an overabundance of Fe compared to He. \textit{Upper right panel:} shows the predicted flux for the $T2$ model, where $T\sim15000\,{\rm K}$. This model is consistent with recent data and naturally explains the expected composition trends, so is used as the fiducial model in additional figures. \textit{Lower left panel:} shows the predicted flux for the $s=2.4$ model. In this model, heavier elements dominate more of the spectrum, so the flux peak reaches relatively higher energies. \textit{Lower right panel:} finally shows the `Amp off' where the non-resonant streaming instability is assumed not to operate (or, equivalently, $\mathcal{A}=1$). This too is consistent with data, but reaches somewhat lower energies and fluxes compared to other models.}
\label{fig:fluxtotals}
\end{figure*}

\begin{equation}\label{flux1}
    d\varepsilon_i(E)=\frac{f_i\epsilon_{\rm CR}dM_{\rm swept}V_{\rm sh}^2/2}{\ln(E_{p,{\rm max}}/E_{p,{\rm min}})}\exp\left(-\ln^2\left(\frac{E}{E_{i,\rm max}}\right)\right),
\end{equation}
where $f_i$ is the fraction of the total flux given to species $i$ (see Table~\ref{tab:cloudyfractions}), assuming a CR production efficiency of $\epsilon_{\rm CR}=0.1$ \cite{Murase:2018okz,Kimura:2018ggg,Kimura:2024lvt}, $dM_{\rm swept}$ is the differential CSM mass swept up (see Eq.~\ref{dmswept}), and $V_{\rm sh}$ is the shock velocity (see Eq.~\ref{shockvelocity}). The denominator is the bolometric correction with minimum and maximum proton energies of $E_{\rm p,min}\sim1\,{\rm GeV}$ and $E_{p\rm,max}$. This results in an escape spectrum of the form above \cite{Zhang:2017moz}, with maximum energy $E_{i\rm,max}$. To provide a time-dependent picture of the CR evolution in each event, we need to integrate this spectrum over the mass swept up, i.e., $\int d\varepsilon_i\propto\int dM_{\rm swept}$.

\begin{figure*}
\centering
\includegraphics[width=0.49\linewidth]{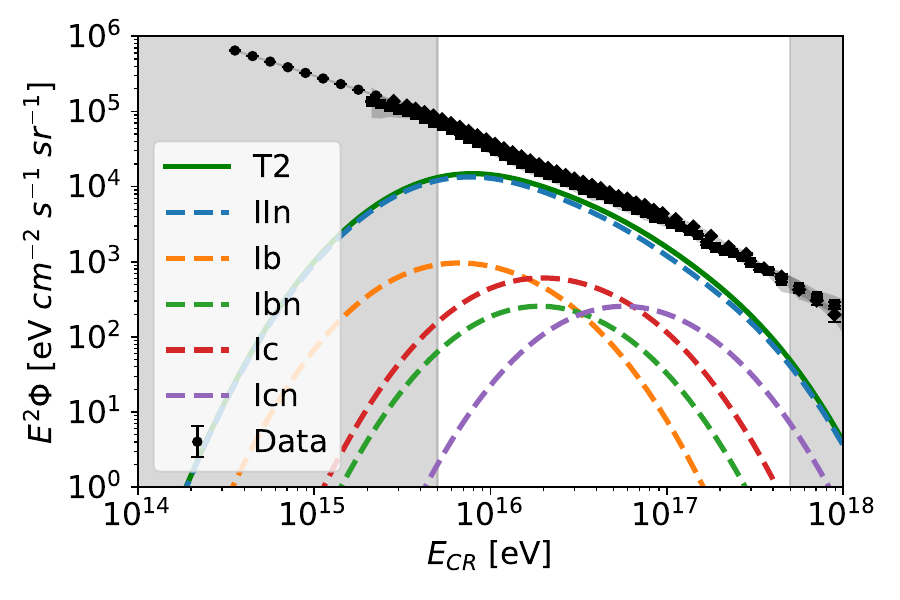}
\includegraphics[width=0.49\linewidth]{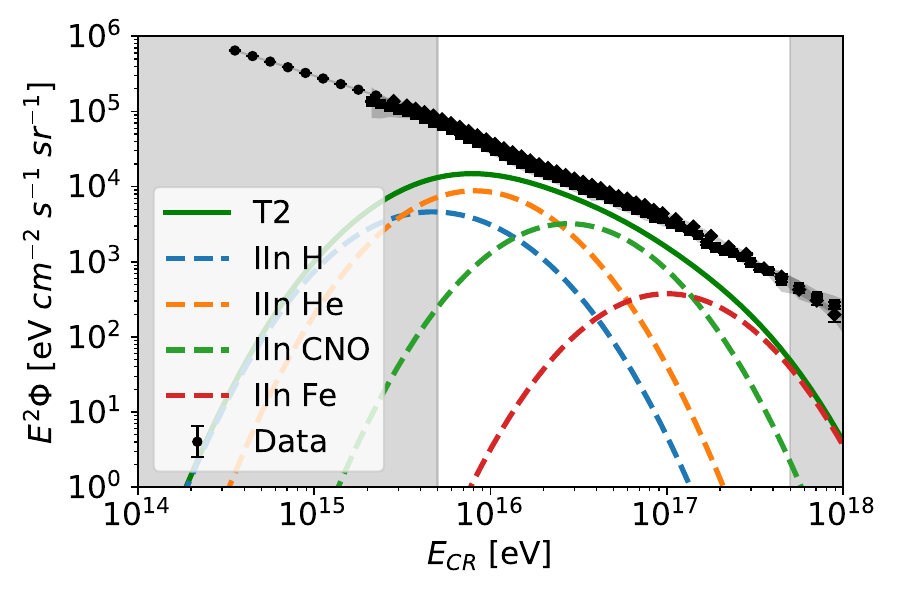}
\caption{Here we break down the components of the $T2$ model's predicted fluxes. \textit{Left panel:} shows the breakdown of supernova types to the overall predicted flux. IIn supernova contribute the most to the predicted flux by $\gtrsim$ an order of magnitude. This is due to the combination of physical parameters (relative to other SN subtypes, the high wind mass-loss rate, low wind velocity, and moderate rate). \textit{Right panel:} shows the breakdown of contributions from H, He, CNO, and Fe nuclei. The dominant component of the predicted flux becomes increasingly heavier with energy because the maximum energies are $A$ and $Z$ dependent.}
\label{fig:fluxcomponents}
\end{figure*}

To estimate the flux at Earth from all ISNe, we first have to estimate the CR production rate, which is simply given by $\int d\varepsilon_i f_{\rm SN}R_{\rm SN}$, where $\mathcal{R}_{\rm SN}\approx10^{-2}\,{\rm yr^{-1}}$ is the rate of local supernovae (estimates vary so this is a representative value \cite{Adams:2013ana,Rozwadowska:2020nab,Quintana25,Ma:2025gbo}). We can then set this production rate equal to the CR escape rate from the CR halo. This escape rate can be expressed as $4\pi M_{\rm gas}E^2\Phi_i/X_{i,\rm esc}$. Here, we assume $M_{\rm gas}=10^{10}\,M_{\odot}$ is the mass of the Milky Way galaxy that CRs travel through, $\Phi_i$ is the CR flux at Earth for species $i$, and $X_{i,\rm esc}$ is the grammage traced by CRs. This quantity is necessary for estimating the effect of partial destruction and diffusion by galactic magnetic fields as CRs propagate from their acceleration sites \cite{Kimura:2018ggg,Murase:2018utn}. The grammage is related to the column density of interstellar material and dependent on the CR rigidity. It can be measured experimentally by observation of the boron-to-carbon flux ratio \cite{Blum:2013zsa} and can be expressed as
\begin{equation}\label{grammage}
    X_{i\rm,esc}=2\,{\rm g~cm^{-2}}
    \begin{cases}
        (\frac{E}{Z_i 250\,{\rm GeV}})^{-0.46}, \hfill E/Z_i < 250\,{\rm GV},\\
        (\frac{E}{Z_i 250\,{\rm GeV}})^{-1/3}, \hfill E/Z_i \geq 250\,{\rm GV}.
    \end{cases}
\end{equation}

Combining Eqs.~\ref{flux1}, the SN rate, and Eq.~\ref{grammage} to transform from $dM_{\rm swept}$ to $dt$, the final steady-state flux is given by
\begin{align}\label{finalflux}
    \nonumber E^2\Phi_i&=\int \frac{f_{\rm SN}\mathcal{R}_{\rm SN}X_{i,\rm esc}}{4\pi M_{\rm gas}} d\varepsilon_i,\\
    \nonumber &=\frac{f_i4\pi C\epsilon_{\rm CR}f_{\rm SN}\mathcal{R}_{\rm SN}X_{i\rm,esc}}{8\pi M_{\rm gas}}\\
    &\times\int_{t_{\rm crit}}^{t_f}\frac{R_{\rm sh}^{2-s}V_{\rm sh}^3}{\ln(E_{p\rm,max}/E_{p\rm,min})}\exp\left(-\ln^2\left(\frac{E}{E_{i\rm,max}}\right)\right)dt.
\end{align}
The total flux for each SN type is given by $\Sigma_iE^2\Phi_i$, summed over each nuclear species $i$. This is repeated for each supernova type and model.

There are several experiments that have recently measured the cosmic-ray flux above the knee as well as the average cosmic-ray composition in this energy range. We compare our predicted models to data from the Telescope Array (TA) low-energy (LE) data set, the IceTop data set, and the more recent LHAASO data set. For the flux comparison, we take data from Refs.~\cite{TelescopeArray:2018bya} (QGSJet II-3 hadronic interaction model), \cite{IceCube:2019hmk} (Sibyll 2.1 model), and \cite{LHAASO:2024knt} (EPOS-LHC model) for the TALE, IceTop, and LHAASO data, respectively. We also include the proton-only/He-only spectra measured by LHAASO \cite{LHAASO:2025byy}/\cite{LHAASO:2025mlf} (EPOS-LHC model). For the composition comparison, we take data from Refs.~\cite{TelescopeArray:2020bfv} (EPOS-LHC model, digitized), \cite{IceCube:2019hmk} (Sibyll 2.1 model), and \cite{LHAASO:2024knt} (EPOS-LHC model), respectively. These data all assume different hadronic interaction models, so there is some additional systematic uncertainty not expressed by the figures (of the order $\lesssim10\%$ \cite{LHAASO:2024knt}). However, these data are still very useful for suggesting which models are reasonable descriptions of the data and generally agree with each other.

\begin{figure*}
\centering
\includegraphics[width=0.49\linewidth]{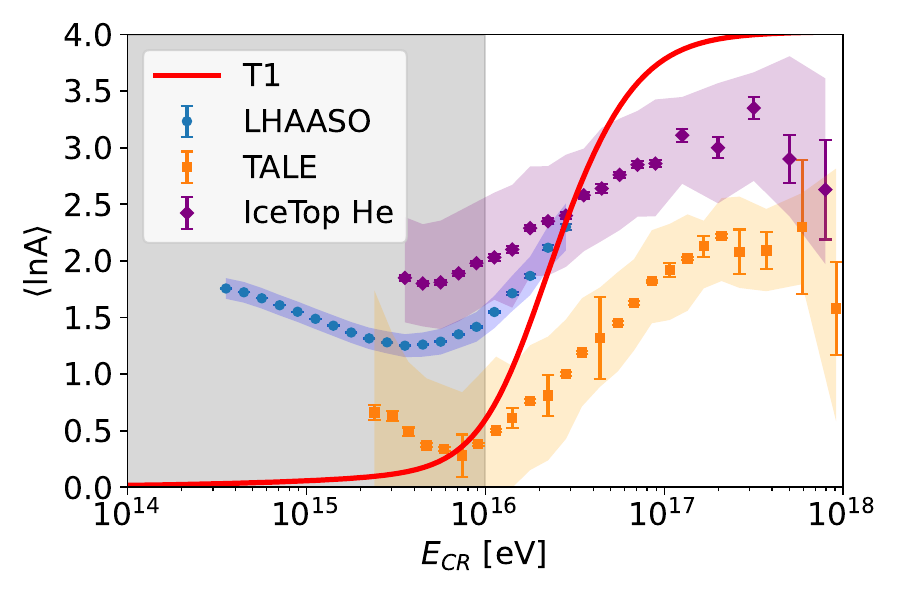}
\includegraphics[width=0.49\linewidth]{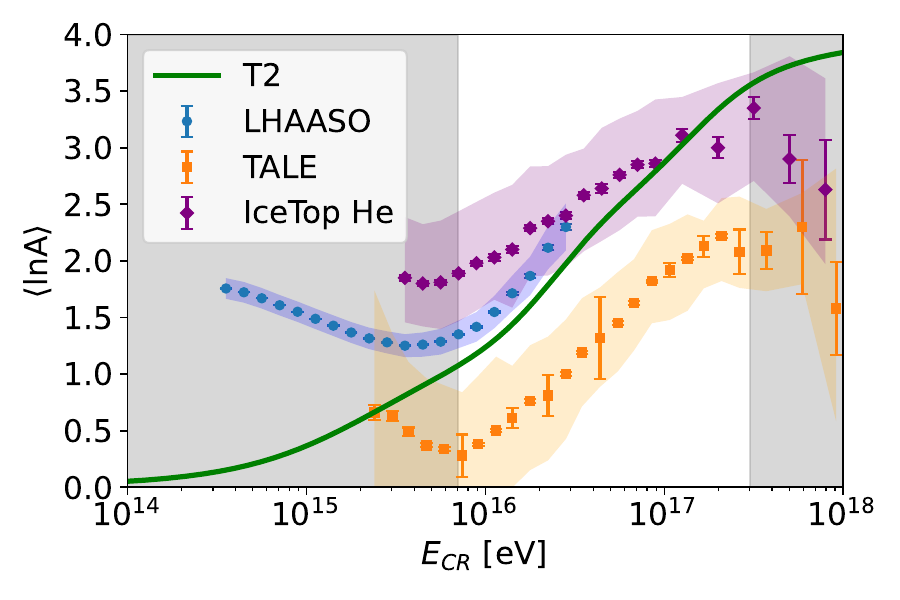}
\includegraphics[width=0.49\linewidth]{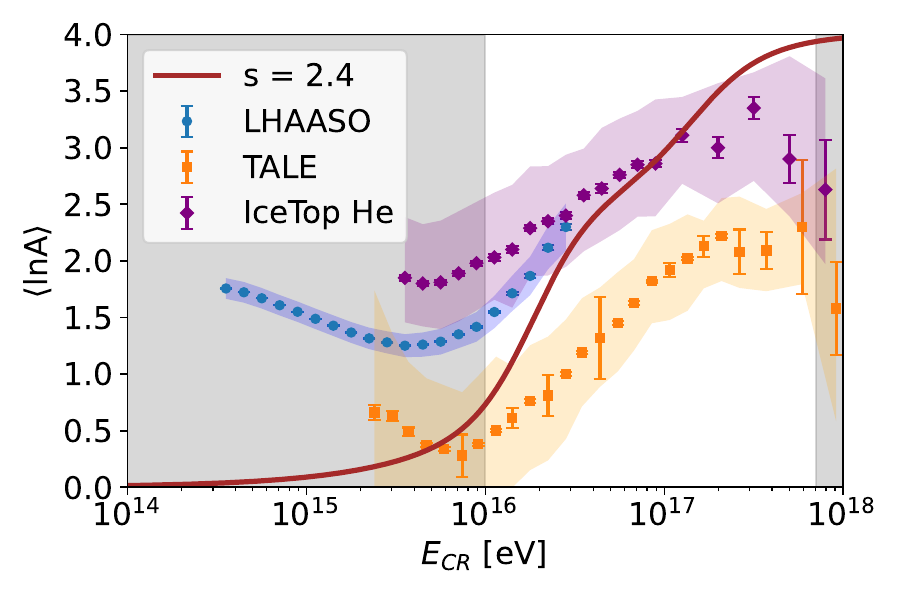}
\includegraphics[width=0.49\linewidth]{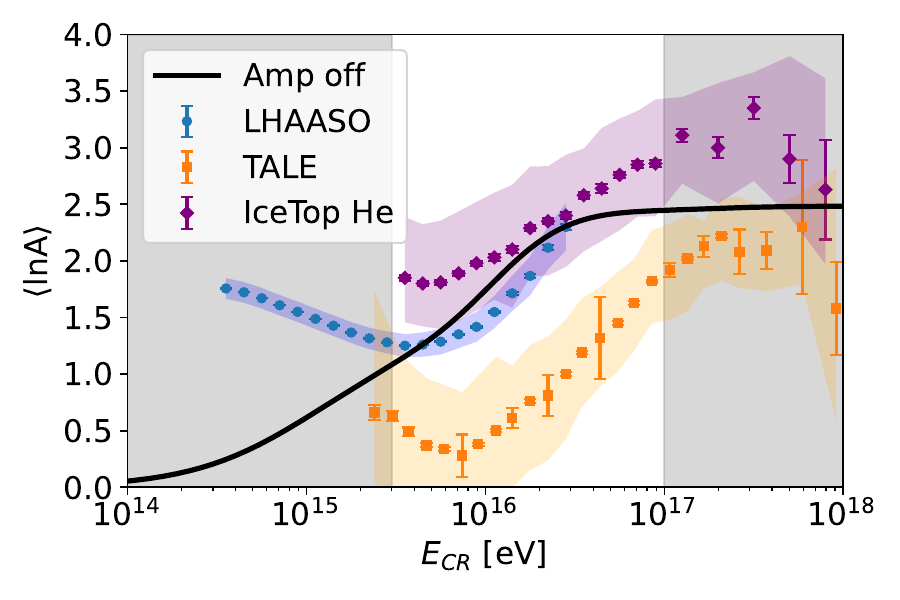}
\caption{The flux-weighted average of the natural log of mass number as a function of energy. Here we show the inferred energy-dependence of composition from the LHAASO, TA, and IceTop experiments compared to that predicted from various models. As with the flux data, these data's statistical errors are shown as vertical bars and systematic errors shown as shaded regions. These figures reveal a potentially large discrepancy between experiments, where the TA observed composition is notably lighter. Since these are all air-shower experiments, the average mass number is an inferred quantity and individual cosmic-rays cannot be detected. \textit{Upper left panel:} shows the average mass number for the $T1$ model, in red. This predicts a significant contribution of protons to the spectrum at $10^{16}\,{\rm eV}$ while overestimating the average mass number at higher energies. \textit{Upper right panel:} shows the average mass number for the $T2$ model, in green, which reasonably agrees with the measured averages between $\sim5\times10^{15}\,{\rm eV}$ and $\sim5\times10^{17}\,{\rm eV}$. \textit{Lower left panel:} shows the $s=2.4$ model, in brown, which shows a similar pattern to the $T1$ result, but may be slightly more consistent with observed averages. Finally, \textit{lower right panel:} shows the `Amp off' model, which similarly is consistent with data, but has a much lower average mass number at the highest energies, compared to other models.}
\label{fig:lnA}
\end{figure*}

To compare our data, the  flux is integrated from $t_{\rm crit}$ (the first time when $E_{\rm max}=E_{\rm esc}$) to $t_f\approx300\,{\rm days}$. In Fig.~\ref{fig:fluxtotals}, we compare the LHAASO, TA, and IceTop data (in blue, orange, and purple, respectively) to the flux predictions for our various models. The statistical error from these experiments is shown with vertical lines, but extremely small, while the systematic errors are given by the shaded regions. Additionally, we show the LHAASO proton-only and He-only spectrum (in gray and black) to check consistency with the composition of CR nuclei. Each model's total flux has a different color, but the p and He components of each are shown in dashed-gray and dashed-black, respectively. The upper left panel of Fig.~\ref{fig:fluxtotals} shows the total flux from our models using the first temperature estimation, $T1$, in green. Although consistent with LHAASO results, the temperature at He escape time ($\sim6600\,{\rm K}$, see Table~\ref{tab:cloudyinputs}) is not high enough to significantly ionize He. This results in a negligible He abundance and an overabundance of heavy elements at the highest energies and, thus, overestimating the total flux above $10^{17}\,{\rm eV}$. In the upper right panel, we compare the data to the $T2$ model, in green. This model is consistent with all data within this energy range and makes up a dominant fraction of the observed flux, especially between $10^{16}\,{\rm eV}$ and $10^{17}\,{\rm eV}$. This model is considered the fiducial model for additional figures for this reason.

Models $T1$ and $T2$ share the same physical parameters but differ in temperature estimation. In the lower left panel of Fig.~\ref{fig:fluxtotals}, we show the $s=2.4$ model which decreases the power-law index from 2.8 to 2.4, but has the same temperature assumption as $T2$. This model is also consistent with data, but because there is a smaller fraction of singly-ionized He, heavier ions dominate the flux and result in a higher energy peak. Finally, in the lower right panel of Fig.~\ref{fig:fluxtotals}, we show the `Amp off' model, which assumes the NRS instability does not operate but otherwise has the same parameters as the $T2$ model. This could occur if, for example, some fiducial parameters like $\xi_{\rm CR}$ or $\overline{\omega}$ are altered. In this case, the results are similar to the $T2$ results, but make up a lower fraction of the total CR flux and reach slightly lower energies.

In Fig.~\ref{fig:fluxcomponents} we also show the various components in the $T2$ model. In the left panel, we show the contribution of the total $T2$ flux from the supernova types we consider. By far, the most important contributor to the cosmic-ray flux is the IIn supernovae. This is primarily because of its comparatively large $\dot{M}$ with smaller $V_w$, while not being too rare. Further, the solar composition assumption helps explain a larger energy window of the cosmic-ray flux. The composition in the stripped-envelope types, meanwhile, are mostly composed of a single nuclei (He or C for Ib/Ibn and Ic/Icn). The Icn supernovae for $\gtrsim10^{17}\,{\rm eV}$, however, do provide a non-negligible fraction of the total flux. In the right panel, the composition breakdown shows how the dominant contribution to the cosmic-ray spectrum becomes increasingly massive.

Next, we compare the flux-weighted average mass number of CR nuclei to the observed values, as a function of energy, in Fig.~\ref{fig:lnA}. It is noteworthy that the total flux from these experiments between $\sim$PeV to EeV agrees well, but the average (natural logarithm) mass number of nuclei shows discrepancies in this energy range. Qualitatively, though, the average mass number increases with energy in this window. Fig.~\ref{fig:lnA} shows the average CR mass numbers for the LHAASO (blue), TA low-energy (TALE, orange), and IceTop high-energy (IceTop HE, purple) experiments. In the upper left panel, we show the results for the $T1$ model. As expected from Fig.~\ref{fig:fluxtotals}, the overabundance of heavy nuclei results in an over-representation in the average mass number figure, and is likely not a good description of the data. The $T2$ model, in the upper right panel, agrees with recent data much better. This highlights the trend of increasingly heavy composition above the knee, up to the second knee. It is heavier on average, though, compared to the TALE results. The lower left panel shows the $s=2.4$ results, which yields qualitatively similar trends to the $T1$ model, but is less divergent from experimental results. Finally, the lower right panel shows the `Amp off' model, which again is overall consistent with the data trends, but shows a lighter composition at the highest energies. For these reasons, we suggest all except the $T1$ model are somewhat reasonable descriptions of the observed CR quantities.

\section{\label{sec:disc}Discussion}

In this work, to show the feasibility of this model, we have to choose parameters for several sources and make assumptions about the CSM structure. Works like Refs.~\cite{Wang:2019bcs,Ransome:2024cza,Wang:2025yem,Dessart:2025lqy} (and the aforementioned sources) use similar parameters to what is used here for modeling ISNe and see similar luminosities, swept up mass, shock radii evolution, etc. However, there seems to be a large diversity in ISN observed properties and, thus, their inferred physical parameters. For ISNe, especially IIn, there is a large variety in the evolution of luminosities, timescales, and energies \cite{Hiramatsu:2024mjz,Salmaso:2024jry,Goto:2025mzs} which can lead to a large variety in inferred mass loss rates, CSM properties, and shock evolution \cite{Kiewe12,Ransome:2024cza} (see also, e.g., \cite{Baer-Way:2024qui,Gangopadhyay:2025gzf}). Additionally, eruptive mass-loss episodes (as opposed to steady winds), lead to different CSM density profiles but still allow for PeV particle acceleration \cite{Brose:2025npd,Brose:2025siv}. Some properties like $t_0$ are more difficult to infer from observations. Further, reducing $\dot{M}$ and $t_0$ together produces a similar result to the fiducial cases shown in this study. Thus, our parameter choices are justified for showing ISNe as a reasonable source of super-knee CRs.

Other studies perform similar analyses of CR acceleration in SNRs and ISNe, including effects of magnetic field amplification \cite{Cristofari:2021hbc}, accounting of various supernova types \cite{Sveshnikova03}, and dense winds from periods of high mass-loss prior to the supernovae \cite{Zirakashvili:2015mua,Murase:2018okz}. Our study combines these features to suggest ISNe, especially IIn, are convincing Super-PeVatrons.

For IIn SNe we assume a `solar' composition, approximated by four species (H, He, CNO, and Fe). We find this is a reasonable first approximation to explain the increasing average mass number of cosmic-rays with energy up to $\sim5\times10^{17}\,{\rm eV}$ because these are some of the most abundant nuclei in the sun. Adding more species, while still considering the preferential injection effect, would provide a more accurate prediction to compare to observations. However, the current disagreement in observed average mass number among experiments prevents this method from being a smoking gun observation.

The recent LHAASO results of the proton-only spectrum \cite{LHAASO:2025byy} and the helium-only spectrum \cite{LHAASO:2025mlf} are useful for constraining our models at energies $\lesssim10^{16}\,{\rm eV}$. Although our models are consistent with these measurements, the proton-only spectrum is useful for excluding some models if IIn supernovae had larger mass-loss rates on average. For example, higher $\dot{M}$ values generally result in a greater flux and higher energies reached, which could overestimate the observed flux at $\sim10^{16}\,{\rm eV}$ for the $T1$ and $s=2.4$ models. These two spectra in combination also help us indirectly understand the temperature dependence in these models. If He is not significantly ionized and remains electrically neutral because of low temperatures, this can lead to larger relative fraction of protons ($f_H$), as is evident in the $T1$ model in Fig.~\ref{fig:fluxtotals}. In the $T2$ and `Amp off' models, He makes up a much larger fraction because it can be more significantly ionized and preferentially injected. Our models do not, however, explain the origin of these spectra, without adding in an additional source. Within this framework, a source that has a higher $s$, lower $\dot{M}$, lower $V_w$, and higher event rate would be required to explain the observed flux, but this is not well motivated by current observations. This source would also need a similar abundance of H and He. Additional, composition-specific, flux measurements would be extremely useful for discriminating models.

\begin{figure}
\centering
\includegraphics[width=\linewidth]{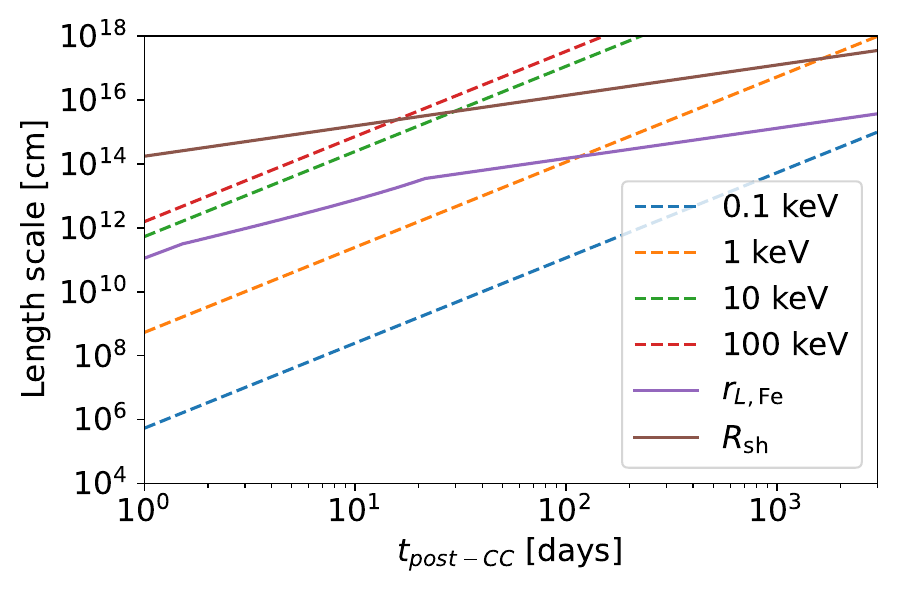}
\caption{The mean free path of X-ray photons (dashed lines) compared to the shock radius, $R_{\rm sh}$, and the Larmor radius, $r_l$, for Fe nuclei. Up to $\sim100\,{\rm days}$, the softer X-ray photons up to $\sim1\,{\rm keV}$ can be reprocessed in the acceleration region of Fe nuclei. Harder X-ray photons, however, have a longer mean free path. The net effect of this may increase the effective temperature that nuclei experience, which affects the ionization state of pre-accelerated nuclei in the CSM and ultimately affects the observed CR composition.}
\label{fig:xraymfp}
\end{figure}

During the shock interaction between the SN ejecta and the CSM, high energy X-rays are created. These X-rays may be partially or fully reprocessed to lower energies, affecting the temperature and, ultimately, the ionization state of the nuclei in the CSM. We assess whether X-rays can be reprocessed by comparing the mean free path of X-ray photons to $R_{\rm sh}$ and the Larmor radius, $r_l$, of nuclei in the acceleration region. For this, we take $E$ here as the maximum energy achieved by accelerated nuclei from Sec.~\ref{sec:acceleration}. Although this is almost rigidity-dependent, for the first $\sim10-20\,{\rm days}$, photodisintegration or spallation are limiting processes, so the mass number plays a role in the maximum energy. For this discussion, we consider Fe ($Z=26$) nuclei (so $E=E_{\rm Fe,max}$), because it has the smallest $r_L$. The mean free path for X-ray photons can be roughly estimated as $\lambda=1/(\kappa_X\rho_{\rm CSM})$ where the X-ray opacity is approximately:
\begin{equation}
    \kappa_X={\rm Max}(10^3E_X^{-3},~\kappa_s)\,{\rm cm^2~g^{-1}},
\end{equation}
where $E_X$ is the X-ray energy in keV. Here, the opacity roughly can be explained by a power-law in energy for soft, low-energy X-rays below $\sim10\,{\rm keV}$ (undergoing mainly photoelectric absorption \cite{Morrison83,Verner96,Wilms:2000ez}), but experience Compton scattering above $\sim10\,{\rm keV}$ at a constant opacity, $\kappa_s=0.34\,{\rm cm^2~g^{-1}}$ \cite{Rybicki79}. Fig.~\ref{fig:xraymfp} shows the result of comparing the $r_l$ of an Fe nucleus at its maximum energy to the X-ray mean free path. This figure suggests that low-energy X-ray photons can easily be reprocessed and thermalized before Fe nuclei are accelerated. However, the high energy photons are not easily reprocessed, so thermalization may be incomplete. This may result in the increased effective temperature that we assume for the $T2$ model. Ref.~\cite{Marcowith:2018ifh} discusses how the luminosity of higher energy X-rays may ionize intermediate-mass elements. Although this, in principle, affects the ionization state and composition of escaping cosmic-rays, a more detailed calculation is outside the scope of this paper.

ISNe are rare compared to the more frequent Ia and II types, so their fraction of the total supernova rate are somewhat uncertain. Recent surveys from Ref.~\cite{Toshikage:2024rzp,Pessi:2025wht}, for example, did not observe Icn supernovae but found IIn and Ibn rates roughly consistent with what was assumed in this work. This would result in a simple rescaling of the overall predicted flux since $\Phi\propto\mathcal{R}_{\rm SN}$. Upcoming surveys like the LSST survey at the Vera C. Rubin Observatory will see $\sim10$ million supernovae over a decade, which can potentially shed light on supernova parameters. However, complementary follow-up observations with spectroscopic information will be needed to classify and learn more about individual supernova types \cite{Ransome:2024cza,Simongini:2025hel}.

Because there are several proposed (Super-)PeVatrons, multimessenger constraints could tell us which contribute most dominantly to the observed CR flux. Ref.~\cite{Kimura:2024lvt} studied the detectability of gamma-rays and neutrinos from SN 2023ixf, the closest type II supernova in recent decades. There, a confined, dense CSM does not result in detectable gamma-rays and neutrinos, in part because SN 2023ixf was still not close enough to Earth. Moreover, 2023ixf-like SNe will not be able to accelerate CRs up to PeV energies. IIn (and other ISNe), though, may be detectable since they harbor physically different CSMs. Regarding SN PeVatrons, Ref.~\cite{Wang:2019bcs} studied the gamma-ray and neutrino emission from the $pp$ interactions of accelerated protons in interacting type II supernovae are detectable up to a few Mpc and can be detected further for denser conditions. Additionally, Refs.~\cite{Murase:2018okz,Murase:2023chr} discuss the multimessenger and multiwavelength prospects from ISNe, including IIn which are promising sources up to several 10s of Mpc, although they are somewhat rare. Considering the composition of accelerated nuclei may also somewhat change the multimessenger signal, since only $pp$ interactions have been considered thus far (see Ref.~\cite{Caprioli:2025lor} for some discussion). Recently, there has been a possible coincidence of neutrino events with a Ibn supernova \cite{Stein:2025uxi} and with two IIn supernovae \cite{Lu:2025jks} (see also \cite{Garrappa:2025ves}), with the possibility of CSM interaction as the source of neutrinos. Gamma-rays and neutrinos are an intriguing way to favor or exclude the models considered here in light of other possible Super-PeVatrons.

\section{\label{sec:summary}Summary}

In this work, we addressed the ability for interacting supernovae to produce high-energy cosmic-rays. We build on a phenomenological model of the shock interaction between SN ejecta and the surrounding CSM, based on physical parameters inferred from observations. We investigate the maximum achievable energies of nuclei accelerated in this interaction, for several ISN types. This maximum energy can be increased if we take into account the magnetic field amplification induced by the streaming of escaping cosmic rays. We also consider the large enhancement to the abundance of partially-ionized heavier species, like C, N, O, and Fe. To do this we model the temperature evolution and calculate the ionization state of nuclei with a state-of-the-art photoionization code, \texttt{Cloudy}. 

Our $T1$ model, where the temperature is proportional to the luminosity of the CSM interaction, does not sufficiently ionize He. This leads to an overprediction of heavier species that overestimate the observed flux at energies higher than $E\gtrsim10^{17}\,{\rm eV}$. However, our other models ($T2$, $s=2.4$, and `Amp off') are consistent with flux measurements and largely consistent with the observed average mass numbers of CRs as well. These models suggest that ISNe - especially IIn - are intriguing candidates of super-knee CRs between $\sim{\rm few}\times10^{15}\,{\rm eV}$ to $\sim{\rm few}\times10^{17}\,{\rm eV}$, able to provide the majority of the flux and explain the increasingly heavy composition up to the second knee.

This work motivates ISNe as viable PeVatron candidates, but additional work is required to confirm them. Radiation-hydrodynamic simulations can help us better understand these objects and connect the physical picture with observations. Clearly detecting gamma-rays and neutrinos coincident with ISNe could confirm them as hadronic accelerators, so calculating the multimessenger signal associated with these events is important. Further, neutrinos are potentially detectable in IceCube and KM3Net if the ISNe are relatively local, even if they are extragalactic \cite{Murase:2017pfe,Petropoulou:2017ymv,Salmaso:2024jry}. This would shed light on whether ISNe are dominant contributors to the high-energy cosmic-ray spectrum or if other astrophysical events are the main PeVatrons.

\begin{acknowledgments}
This work is partly supported by KAKENHI Nos. 22K14028, 21H04487 (S.S.K.), 23H04899 (S.S.K. \& K.K.), and JP22H00130, JP24K00668, JP25K00021 (K.K.). 
S.S.K. acknowledges the support by the Tohoku Initiative for Fostering Global Researchers for Interdisciplinary Sciences (TI-FRIS) of MEXT's Strategic Professional Development Program for Young Researchers.
\end{acknowledgments}

\newpage

\appendix

\section{\label{sec:icetopresults} Composition comparison to IceTop}

Here we also compare the composition breakdown of the flux from our four models to IceTop results. Although these comparisons yield similar qualitative results to those of Fig.~\ref{fig:fluxtotals}, these break down the inferred composition at higher energies. Note that, for $E_{\rm CR}\lesssim10^{16}\,{\rm eV}$, each of the four components have a similar level of flux, but at higher energies, Fe and O are increasingly abundant. 

As in previous discussion, the $T1$ model should be largely disfavored due to a negligible He contribution to the observed flux and an overprediction of Fe nuclei to the observed flux. The $T2$, on the other hand, is consistent with the observed data, but cannot explain the entire observed flux. The CNO component of the $s=2.4$ model may be marginally consistent with the observed flux of O nuclei, but is able to provide the Fe flux at high energies. Finally, the Amp off model results are similar to $T2$, but has overall a lower flux and smaller energies reached. All these data support that ISNe can be a major contributor to the super-PeV CR flux, but additional sources are required at energies $\lesssim{\rm few}\times10^{15}\,{\rm eV}$ and $\gtrsim{\rm few}\times10^{17}\,{\rm eV}$.

\begin{figure*}
\centering
\includegraphics[width=0.49\linewidth]{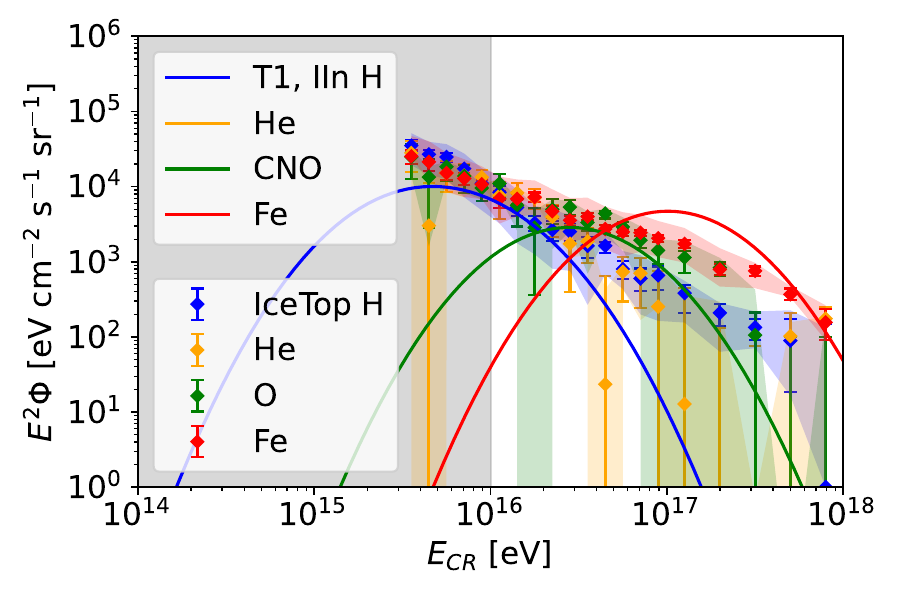}
\includegraphics[width=0.49\linewidth]{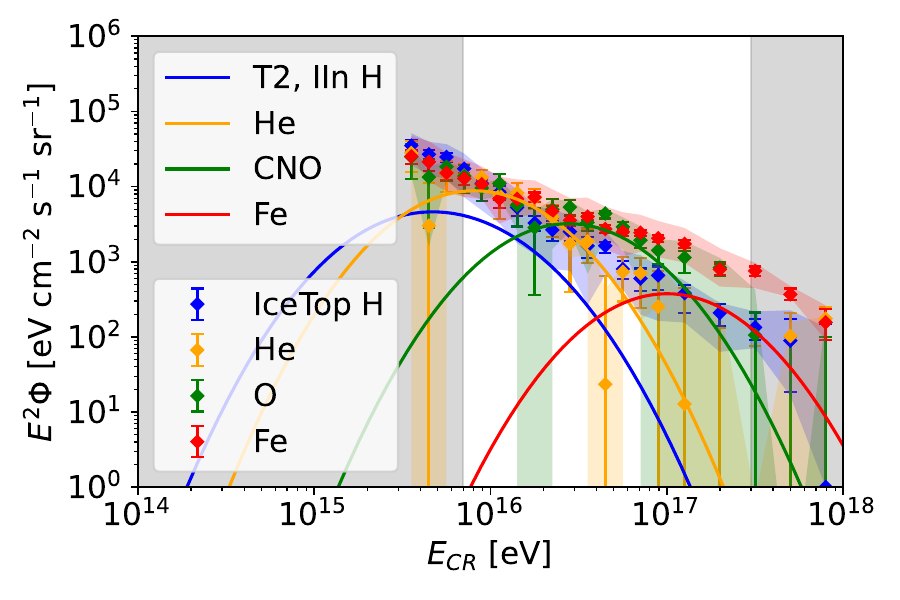}
\includegraphics[width=0.49\linewidth]{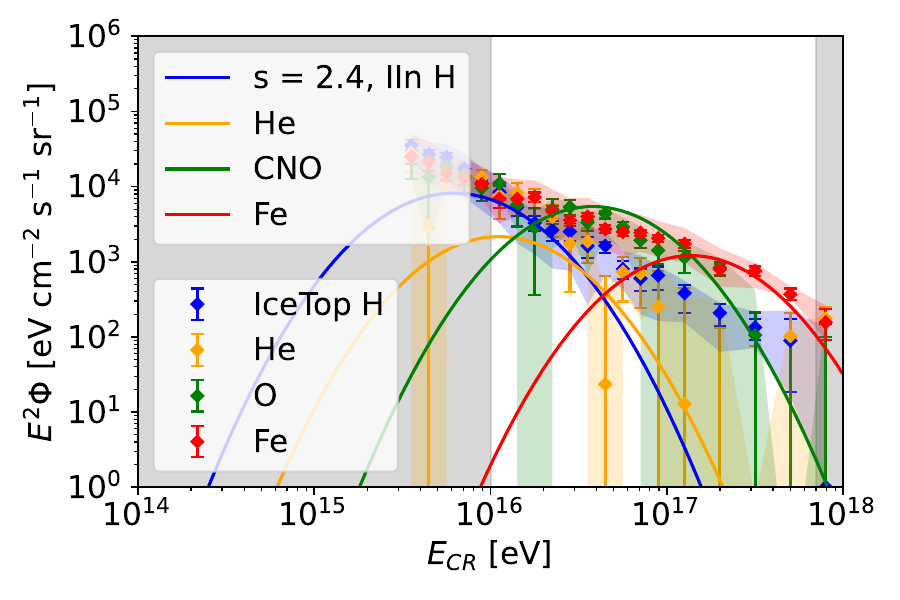}
\includegraphics[width=0.49\linewidth]{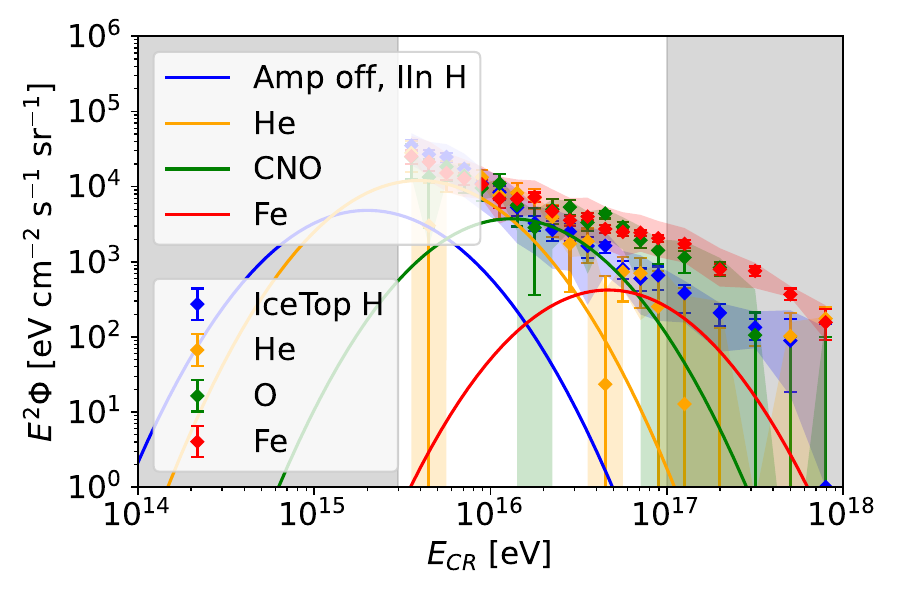}
\caption{Breakdown of the flux from H, He, CNO, and Fe compared to the broken flux from IceTop data. Note that each component has similar flux below $\sim10^{16}\,{\rm eV}$, but becomes relatively heavier at higher energies. The models shown in each panel are the $T1$, $T2$, $s=2.4$, and Amp off models also shown in Figs.~\ref{fig:fluxtotals} and \ref{fig:lnA}.}
\label{fig:icetopcomp}
\end{figure*}

\bibliography{main}

\end{document}